\documentclass[11pt,a4paper]{article}
\usepackage{jheppub}
\usepackage{amssymb,amsmath,amsfonts,mathrsfs}
\usepackage{graphicx}
\usepackage[export]{adjustbox}

\usepackage[dvipsnames]{xcolor}
\usepackage{verbatim}
\usepackage{epsfig}
\usepackage{color}
\usepackage{multirow}
\usepackage{comment}
\usepackage{datetime}
\usepackage{slashed}
\usepackage{framed}
\usepackage{longtable}
\usepackage[mathscr]{euscript}
\usepackage{cancel}
\usepackage{tensor}
 \usepackage{mathtools}
  \usepackage{caption}
  \usepackage{subcaption}
 \usepackage[multiple]{footmisc}
 \usepackage{pgfplots}
\usepackage[applemac]{inputenc}
\usepackage{tikz}
\usepackage{placeins}
\usetikzlibrary{decorations.markings}

\def\be{\begin{equation}}
\def\ee{\end{equation}}
\def\bea{\begin{align}}
\def\eea{\end{align}}

\title{A Rutherford-like formula  for scattering off Kerr-Newman BHs and subleading corrections}

\author[a]{Massimo Bianchi,}
\author[b]{Claudio Gambino,}
\author[b]{Fabio Riccioni\,}

\affiliation[a]{Dipartimento di Fisica,  Universit\`a di Roma ``Tor Vergata"  \& Sezione INFN Roma2, Via della Ricerca Scientifica 1, 00133, Roma, Italy}
\affiliation[b]{Dipartimento di Fisica,  Universit\`a di Roma ``La Sapienza"  \& Sezione INFN Roma1, Piazzale Aldo Moro, 00184, Roma, Italy}

\abstract{By exploiting the Kerr-Schild gauge, we study the scattering of a massive (charged) scalar off a Kerr-Newman black hole. In this gauge, the interactions between the probe and the target involve only tri-linear vertices. We manage to write down the tree-level scattering amplitudes in analytic form, from which we can construct an expression for  the eikonal phase which is exact in the spin of the black hole at arbitrary order in the Post-Minkowskian expansion. We compute the classical contribution to the cross-section and deflection angle at leading order  for a Kerr black hole for arbitrary orientation of the spin.  Finally, we test our method by reproducing the classical  amplitude for a Schwarzschild black hole at second Post-Minkowskian order  and outline how to extend the analysis to the Kerr-Newman case.

}

\begin{document}
\maketitle
\flushbottom 

\section{Introduction}
The  formula for the scattering of $\alpha$ particles of energy $E = mv^2/2$ off nuclei
\be
{d\sigma \over d\Omega}\Big|_{Ruth.} = {Z_\alpha^2 Z_{Au}^2 e^4 \over 16 E^2\sin^4{\vartheta\over 2}}\ ,
\ee 
with $Z_{\alpha}=2$ (Helium) and $Z_{Au}=79$ ({\it Aurum}, aka gold),  is the corner-stone of any theory of fundamental interactions \cite{Rutherford:1911zz}. Its relativistic generalization to spin 1/2 particles (e.g. electrons) by Mott \cite{Mott} 
represents the simplest non-trivial application of QED.

The generalization to the scattering off Black-Holes (BHs) in General Relativity has attracted a lot of attention over the years, following the seminal work of \cite{matzner1968scattering}. 
While for non-rotating (Schwarzschild and Reissner-Nordstr\"om) BHs radial and (polar) angular motion completely decouple \cite{PhysRev.108.1063, Chandrasekhar:1985kt} so that, without loss of generality, one can always work on the `equatorial' plane, for rotating BHs (Kerr and Kerr-Newman) separation of the dynamics is possible, as originally observed by Carter \cite{PhysRev.174.1559}, but the separation constant depends on the BH spin and the energy of the probe \cite{Teukolsky:1973ha,Press:1973zz,Teukolsky:1974yv}. 
For this reason, Rutherford-like formulae for scattering off Schwarzschild BHs can be relatively easily written down, while for Kerr and Kerr-Newmann BHs only partial results have been obtained for special kinematical configurations, {\it i.e.} on-axis incidence, in Black Hole Perturbation Theory (BHPT) \cite{Doran:2001ag, Glampedakis:2001cx,Dolan:2008kf, Hoogeveen:2023bqa}. 

Recently, the study of the two-body problem in General Relativity received renewed interest due to its relation to the 
analysis of the inspiral phase of BH mergers and other extreme processes \cite{Buonanno:2022pgc, Bjerrum-Bohr:2022ows}. In this context, the modern approach is to treat General Relativity as an effective field theory \cite{Donoghue:1994dn}, and consider extended objects like black holes as spinning point particles, whose interactions are then  computed using scattering amplitude techniques in quantum field theory. Within this framework, it is possible to derive both metrics \cite{Donoghue:2001qc, Bjerrum-Bohr:2002fji, Jakobsen:2020ksu, Mougiakakos:2020laz, DOnofrio:2022cvn,Gambino:2022kvb} and gravitational observables \cite{Kosower:2018adc, Cristofoli:2021vyo, Kalin:2019rwq, Kalin:2019inp, Kalin:2020mvi, Bjerrum-Bohr:2013bxa,
Neill:2013wsa}. As it is natural from a quantum field theory perspective, the observables are expanded in a Post-Minkowskian (PM) series, {\it i.e.} an expansion in the coupling $G$, corresponding to a loop expansion in the scattering amplitude. Moreover, recent developments allow to extract the classical contribution before computing  the full amplitude at each order in the PM expansion, remarkably simplifying the computations involved \cite{Bjerrum-Bohr:2018xdl, Bjerrum-Bohr:2021vuf, Guevara:2017csg, Brandhuber:2021eyq, Brandhuber:2021kpo}.

A systematic approach to derive the classical contribution to the scattering process is based on the eikonal exponentiation \cite{Amati:1987wq, Amati:1987uf, Amati:1990xe, Verlinde:1991iu, Kabat:1992tb, Levy:1969cr, AccettulliHuber:2020oou, Cristofoli:2020uzm}, using the principle that in the classical limit the S-matrix is $e^{2i\delta}$, where the eikonal phase $\delta$ plays the role of the action, which is large in $\hbar$  units. In the case of the scattering of non-spinning BHs, enormous progress has been achieved in the analysis of the conservative dynamics \cite{Brandhuber:2023hhy, DiVecchia:2022nna, Bern:2021yeh, Jakobsen:2023ndj, Bern:2019nnu, Bern:2019crd, Cheung:2020gyp, Kalin:2020fhe, Dlapa:2021npj, Dlapa:2021vgp, Bern:2021dqo, Brandhuber:2021eyq}, and recently radiation effects have been included \cite{Dlapa:2022lmu, Goldberger:2009qd, Herrmann:2021tct, Jakobsen:2022zsx, Jakobsen:2022fcj, Bini:2022wrq, Bini:2022enm, Bini:2021gat, Damour:2020tta}, as well as `tail effects' \cite{Bern:2022kto, Goldberger:2004jt, Porto:2016pyg, Kalin:2020lmz, Bern:2021yeh, Bini:2020flp, Cheung:2020sdj}.  Much less is known in the case of the scattering of spinning objects. In this context, following the work of \cite{Holstein:2008sx, Arkani-Hamed:2017jhn, Chung:2018kqs} a lot of effort has been devoted to the study of higher-spin scattering amplitudes \cite{Aoude:2022thd, Bern:2022kto, Georgoudis:2023lgf, Guevara:2018wpp, Arkani-Hamed:2019ymq, Moynihan:2019bor, Bern:2020buy, Jakobsen:2023ndj}, while in \cite{Bautista:2021wfy,Bautista:2022wjf, Guevara:2019fsj}, following the idea of \cite{Vines:2017hyw}, the dynamics of such system was derived from the knowledge of the exact Kerr metric. This allows to write down a tri-linear vertex describing the interaction of the probe with the BH background, giving the amplitude at 1PM, while additional vertices are needed to compute the amplitude at higher PM orders.  

In this paper we study the scattering of a massive (charged) scalar off a Kerr-Newman (KN) BH exploiting the remarkable properties of the Kerr-Schild (KS) gauge. Indeed, in this gauge the first-order expansion in $G$ of the metric is exact and the interaction between the scalar and the background metric is completely described by a single tri-linear vertex. This holds true also for the interaction of a charged scalar with the electric potential generated by the charge of the BH. Besides, another special feature of this gauge is that it allows us to determine the exact expression of the metric and electric potential in momentum space, from which we write down the tree-level scattering amplitude of a massive scalar probe off a KN background in a compact analytic form.
From the amplitude we derive the tree-level cross-section for a massive particle off a Kerr BH for arbitrary orientation of the angular momentum of the BH with respect to the angular momentum of the probe.

Performing a Fourier transform to impact parameter space, we then show how the tree-level scattering amplitude alone allows us to determine  not only the leading contribution to the scattering angle, but also the  sub-leading corrections, using the eikonal expansion. Indeed, the absence of higher-order interaction vertices between scalars and gravitons or photons in the KS gauge  implies that all the diagrams that contribute to the scattering amplitude at higher orders in $G$ in the classical limit are `comb-like' diagrams as in Fig.~\ref{L_loop_Amp}. 
We first  determine the leading eikonal for a probe scattering off KN BHs and the corresponding deflection angle.  We then study the  2PM amplitude in the simpler case of a Schwarzschild black hole. Apart from recovering the standard hyper-classical term which reproduces the exponentiation, we show that the classical term arises in this approach from the presence in the amplitude of `off-shell' terms, whose appearance is a trademark of the KS gauge.

It is worth commenting on the relation of our work with the existing literature. First of all, the charged sector of our amplitude fully agrees with the analysis of \cite{Chung:2019yfs}. Moreover, as far as the leading eikonal is concerned, our results coincide for Kerr BHs with the ones in \cite{Bautista:2021wfy}, where the amplitude is computed using an on-shell formalism and the local terms, that are not relevant in the classical limit, are dropped from the start, following the procedure outlined in \cite{Guevara:2017csg}. However, to the best of our knowledge  there are no similar results for the terms proportional to the square of the charge in the KN case. Concerning subleading corrections, it is remarkable that the off-shell terms in the comb-like amplitudes reproduce the classical contribution associated to contact terms, {\it i.e.} vertices with more than one graviton attached. As an additional remark, we point out that while the literature is mainly focused on the neutral case, studying the scattering of charged objects may find application in the context of dark-photon scenari \cite{Cardoso:2016olt}. 

The paper is organised as follows. In section \ref{Sec:FT} we review the Kerr-Newman solution in KS gauge, and  we compute its Fourier Transform (FT) to momentum space, which is exact, although at first order in $G$, thanks to the miraculous properties of the gauge. In section \ref{Sec:Cross_Section} we derive the 1PM scattering amplitudes between a massive charged scalar probe and a KN background. We then restrict to the Kerr case and use the relevant amplitude to determine the tree-level cross-section for arbitrary spin orientation. In section \ref{Sec:Eik_Expansion} we show how the eikonal expansion is performed in this gauge, and in particular how off-shell terms in the amplitude conspire to reproduce contact terms. In section \ref{Sec:Leading_Eik_Kerr} we determine the leading eikonal phase and deflection angle. In particular, for Kerr BHs we derive an expression  which is formally exact in the spin of the black hole  and coincides with  \cite{Guevara:2018wpp}, while for the charge contribution we write down the eikonal phase in an integral form, from which every term in the expansion in the spin can easily be derived. In section \ref{subleadingeikonalsection} we study the sub-leading eikonal corrections. We first consider the Schwarzschild case, and show how our method reproduces the results known in the literature, and we then comment on how this can be extended to KN. We also briefly comment on the graviton self-interaction contribution. Finally, section \ref{Sec:Conclusions} contains our conclusions.

\section{Kerr-Newman solution in Kerr-Schild gauge and its Fourier Transform}\label{Sec:FT}
In this section we show that working in the KS gauge allows us to compute exactly the FT of the metric describing the KN solution. We work in the  $(+---)$ signature and in natural units $c=\hbar =1$, keeping the dependence on $G$ explicit. 
Using oblate spheroidal (OS) coordinates \`a la Boyer-Lindquist, whereby
\be 
x\pm i y = \sqrt{r^2+a^2} \sin\vartheta \ e^{ \pm i\varphi}\quad \text{and}\quad   z= r\cos\vartheta \ ,
\ee
such that 
\be
{x^2 + y^2 \over r^2 + a^2} + {z^2\over r^2} =1\ ,
\ee
the KN metric in KS gauge reads \cite{Debney:1969zz,Adamo:2014baa}
\be
g_{\mu\nu} =\eta_{\mu\nu} + h_{\mu\nu} =\eta_{\mu\nu} + \Phi_G K_{\mu} K_{\nu}\ , \label{KerrSchieldx}
\ee
where $\Phi_G$ is the `gravitational' potential
\be
\Phi_G= - G{2M r - Q^2 \over r^2 + a^2 \cos^2\vartheta}\ ,\label{gravitationalpotential}
\ee
with $M$ the mass of the BH, $a = J/M\le M$ its angular momentum per unit mass (oriented along the $z$-axis as usual), $Q$ its electric charge and $K_\mu$ the null vector  
\be
K_\mu = \left( 1, {rx+ay \over r^2 + a^2}, {ry-ax \over r^2 + a^2}, {z\over r}\right)\ .
\ee
Notice that $K$ is null with respect to both $\eta$ and $g$ so much so that the inverse metric reads
\be
g^{\mu\nu} = \eta^{\mu\nu}+h^{\mu\nu}=\eta^{\mu\nu} - \Phi_G K^{\mu} K^{\nu}\ ,
\ee 
with $K^\mu \equiv g^{\mu\nu} K_\nu = \eta^{\mu\nu} K_\nu$. 
Moreover $\det(g) = \det(\eta)=-1$ since $\text{Tr}(h)=\eta^{\mu\nu} h_{\mu\nu} =h= 0$ and $h_{\mu\nu} h^{\nu\lambda} =0$ so that $h^n=0$ for $n>1$. 

The `electromagnetic' 4-potential  generated by the BH charge \cite{Adamo:2014baa} is given by 
\be
A_\mu= V_{A} {K}_\mu\ ,\label{KerrSchieldAx}
\ee
where $K_\mu$ is the same null vector as in the metric and $V_A$ is the `electric' potential 
\be
V_{A} = {Qr \over r^2+a^2\cos^2\vartheta}\ .
\ee
Notice that the relation between the potential and the metric in KS gauge resembles the double copy construction of \cite{Monteiro:2014cda}.

For later purposes, we need the FT of the gravitational field ${h}_{\mu\nu}$ and 4-potential $A_\mu$ in the KS gauge. 
Since the metric is stationary the FT can be written as\footnote{We denote with $\hat{h}$ the FT to the 4-dimensional momentum space and with $\tilde{h}$ the one to the space-like momentum.}
\be
\hat{h}^{KN}_{\mu\nu}(q) = 2 \pi \delta(q_0) \tilde{h}^{KN}_{\mu\nu}(\vec{q}\,)= 2 \pi \delta(q_0) \int d^3 \vec x e^{-i\vec q\cdot \vec x} \Phi_G(\vec{x}\,) K_{\mu}(\vec{x}\,) K_{\nu}(\vec{x}\,)\ ,\label{FTwithdelta}
\ee
where the $2 \pi \delta(q_0)$  factor in front is due to the fact that the space-time is stationary. 
In OS coordinates we have 
\be
d^3\vec x = dxdydz= (r^2+a^2\cos^2\vartheta) \sin\vartheta dr  d\vartheta d\varphi \ ,
\ee
and the factor $(r^2+a^2\cos^2\vartheta)$ exactly cancels the denominator in $\Phi_G(r,\vartheta)$ (independent of $\varphi$, thanks to axial symmetry) and we end up with 
\be
\tilde{h}^{KN}_{\mu\nu}(\vec{q}\,) = -G \int dr (2Mr-Q^2) \sin\vartheta d\vartheta d\varphi e^{-i\vec{q}\cdot \vec{x}}  K_{\mu}(r,\vartheta) K_{\nu}(r,\vartheta)\ ,
\ee
where 
\be \vec{q}\cdot \vec{x} = (q_x \cos\varphi + q_y \sin\varphi) \sin(\vartheta)\sqrt{r^2+a^2} + q_z r\cos\vartheta \ .
\ee
Replacing then $\vec{x}$ with $i \partial/\partial\vec{q}$, while keeping the dependence on $r$, gives
\be
\tilde{h}^{KN}_{\mu\nu}(\vec{q}\,) = -G \int dr (2Mr-Q^2) \sin\vartheta d\vartheta d\varphi  
\hat K_\mu (r, \vec{x}=i \partial_{\vec{q}}) \hat K_\nu (r, \vec{x}=i \partial_{\vec{q}}) e^{-i\vec{q}\cdot \vec{x}} \ ,
\ee
where
\be
\hat K_\mu\left(r, \vec{x}=i {\partial_{\vec{q}}}\right)= \left( 1, i{r\partial_{q_x}+a\partial_{q_y} \over r^2 + a^2}, i{r\partial_{q_y}-a\partial_{q_x} \over r^2 + a^2}, i{\partial_{q_z}\over r}\right)
\ee
is a differential operator in $\vec{q}$ space.
Now the angular measure $\sin\vartheta d\vartheta d\varphi= d\Omega$ is the rotation invariant measure on the 2-sphere and 
$\vec{q}\cdot \vec{x}$ can be written as 
\be
(q_x \cos\varphi + q_y \sin\varphi) \sin (\vartheta)\sqrt{r^2+a^2} + q_z r\cos\vartheta = \vec{u}{\cdot}\vec{n}\ ,
\ee
with 
\be 
\vec{n} = (\sin\vartheta\cos\varphi , \sin\vartheta\sin\varphi, \cos\vartheta)
\ee
the standard unit vector on the 2-sphere and 
\be
\vec{u} = (q_x \sqrt{r^2+a^2},  q_y \sqrt{r^2+a^2}, q_z r)\ ,
\ee
whose length is 
\be
u=|\vec{u}| = \sqrt{r^2 q^2 + a^2 q_\perp^2}
\ee
with 
\be
{{|\vec{q}\,|}}^2 = q_x^2 + q_y^2 + q_z^2 = q_\perp^2 + q_z^2\ .
\ee
We can thus perform the elementary integral over the solid angle
\be
\int d\Omega e^{-i \vec{u}{\cdot}\vec{n}} = 4\pi {\sin{u} \over u} = 4\pi j_0(u)
\ee
and get 
\be
\tilde{h}^{KN}_{\mu\nu}(\vec{q}\,) = -4 \pi G \int_0^\infty dr (2Mr-Q^2) \hat K_\mu (r, \vec{x}=i \partial_{\vec{q}}) \hat K_\nu (r, \vec{x}=i \partial_{\vec{q}}) {j_0(u(r, \vec q\, ))} \ , \label{FTmetric}
\ee
where we denote with $j_n$ the spherical Bessel functions, that for $n=0, 1$ read
\be
j_0(x)=\frac{\sin x}{x}\ , \qquad j_1(x)=\frac{\sin x}{x^2}-\frac{\cos x}{x}\ .
\ee
The integral over the radial direction is begging to be performed changing variable from $r$ to $u$, whereby
\be
{{|\vec{q}\,|}}^2 r^2 = u^2- a^2 q_\perp^2 \ ,\qquad {{|\vec{q}\,|}}^2 (r^2+a^2) = u^2+ a^2 q_z^2 \quad {\rm and} \quad  rdr = {udu \over {{|\vec{q}\,|}}^2} \ .
\ee
Keeping in mind that $u$ depends on $\vec{q}$, it is useful to consider how $\hat K_\mu$ acts on a function $F(u)$
\be
\hat K_\mu(u, i\partial_{\vec q})F(u)=(1, i(r(u)q_x+aq_y), i(r(u)q_y-aq_x), ir(u)q_z)\frac{1}{u}\frac{d}{du}F(u)\ .
\ee
The master integrals which are needed in order to express the results in an analytic form are
\be
C_n=\int_{q_\perp a}^{+\infty}du\frac{u^{1-n} j_{n}(u)}{\sqrt{u^2-q_\perp ^2 a^2}}=\frac{\pi}{2}\frac{J_n(q_\perp a)}{(q_\perp a)^n}  \quad {\rm with} \quad n=0,1,2\ ,\label{masterintegral}
\ee
where $J_n$ are the Bessel functions of the first kind, defined in terms of one of their representation as
\be
J_n(x)=\sum_{m=0}^{+\infty}\frac{(-1)^m}{m!\, \Gamma(m+n+1)}\left(\frac{x}{2}\right)^{2m+n}\ .
\ee

Let us focus first  on the case of Kerr BHs ($Q=0$).
Dropping the overall factor $-8\pi G M$ common to all $\tilde{h}_{\mu\nu}(q)$ (which will be reinserted back in the computation of the amplitude) the FT assumes the form
\be
\tilde{h}_{\mu\nu}(\vec{q}\,) = \int_{q_\perp a}^\infty {udu \over {{|\vec{q}\,|}}^2}  \hat {K}_\mu (u, \vec{x}=i \partial_{\vec{q}}) \hat {K}_\nu (u, \vec{x}=i \partial_{\vec{q}})j_0(u)\ .
\ee
After using eq. \eqref{masterintegral} and regulating with $e^{-\varepsilon u}$ when necessary, 
one gets 
\be
\begin{aligned}
 \tilde{h}_{00}(\vec q\,) & = {1\over {{|\vec{q}\,|}}^2} \cos|\vec{a}{\times}\vec{q}\,|\ , \\  
 \tilde{h}_{0i}(\vec q\,) & =-i\frac{q_i}{{{|\vec{q}\,|}}^3}\frac{\pi}{2}J_0(|\vec{a}{\times}\vec{q}\,|)+i\frac{(\vec{a}\times \vec{q}\,)_i}{{{|\vec{q}\,|}}^2}j_0(|\vec{a}{\times}\vec{q}\,|)\ ,\\
\tilde{h}_{ij}(\vec q\,)&=\frac{j_0(|\vec{a}{\times}\vec{q}\,|)}{{{|\vec{q}\,|}}^2}\left(\delta_{ij}-2\frac{q_iq_j}{{{|\vec{q}\,|}}^2}\right)+\frac{1}{{{|\vec{q}\,|}}^3}\frac{\pi}{2}\frac{J_1(|\vec{a}{\times}\vec{q}\,|)}{|\vec{a}{\times}\vec{q}\,|}\Bigl(q_i (\vec{a}\times \vec{q}\,)_j+q_j (\vec{a}{\times}\vec{q}\,)_i\Bigl)\\ 
&-\frac{1}{{{|\vec{q}\,|}}^2}\frac{j_1(|\vec{a}{\times}\vec{q}\,|)}{|\vec{a}{\times}\vec{q}\,|}(\vec{a}\times\vec{q}\,)_i(\vec{a}\times\vec{q}\,)_j \ . \label{htilde}
\end{aligned}
\ee
In order to write down the result in this compact form, one has simply to observe that
\be
(-a q_y, a q_x, 0)=(\vec a \times \vec q\,) \quad {\rm and} \quad 
q_\perp a = |\vec{a}{\times}\vec{q}\,| \ . 
\ee
Besides, in the computations it is helpful to notice that the $y$-components follow from the $x$-components after replacing $q_x\rightarrow q_y$ and $q_y\rightarrow -q_x$.
As a `sanity check' one can easily verify  that $\eta^{\mu\nu} \tilde{h}_{\mu\nu} =0$ following from $K_\mu K^\mu=0$: the hallmark of  KS gauge.

For charged KN BHs, the gravitational potential $\Phi_G$ in \eqref{gravitationalpotential} receives an additional contribution proportional to $Q^2$.  Plugging this into eq. \eqref{FTmetric}, we notice that the dependence on $r$ is different with respect to previous terms.
Neglecting the overall common factor ${4\pi G Q^2}$, 
the relevant shifts are given by 
\be
\Delta\tilde{h}_{\mu\nu}(\vec{q}\,) = \int_{q_\perp a}^\infty {udu \over {{|\vec{q}\,|}}^2} \frac{1}{r(u)} \hat {K}_\mu (u, \vec{x}=i \partial_{\vec{q}}) \hat {K}_\nu (u, \vec{x}=i \partial_{\vec{q}})j_0(u)\ .
\ee
The integrals are computed as before using the master integrals in eq. \eqref{masterintegral}, and we find

\begin{align}
\Delta\tilde{h}_{00}(\vec{q}\,) &=\frac{1}{{{|\vec{q}\,|}}}\frac{\pi}{2}J_0(|\vec a \times \vec q\,|)\ , \nonumber \\ 
\Delta \tilde{h}_{0i}(\vec q\,) & =-i\frac{q_i}{{{|\vec{q}\,|}}^2}j_0(|\vec a \times \vec q\, |)+i\frac{(\vec a \times \vec q\,)_i}{{{|\vec{q}\,|}}}\frac{\pi}{2}\frac{J_1(|\vec a \times \vec q\,|)}{|\vec a \times \vec q\,|} \ , \label{Deltahtilde}\\
\Delta \tilde{h}_{ij}(\vec q\,)&=\frac{1}{{{|\vec{q}\,|}}}\frac{\pi}{2}\frac{J_1(|\vec a \times \vec q\,|)}{|\vec a \times \vec q\,|}\left(\delta_{ij}-\frac{q_iq_j}{{{|\vec{q}\,|}}^2}\right)+\frac{1}{{{|\vec{q}\,|}}^2}\frac{j_1(|\vec a \times \vec q\,|)}{|\vec a \times \vec q\,|}\Biggl(q_i(\vec a\times \vec q\,)_j+q_j(\vec a\times \vec q\,)_i\Biggl)\nonumber\\ 
&-\frac{1}{{{|\vec{q}\,|}}}\frac{\pi}{2}\frac{J_2(|\vec a \times \vec q\,|)}{|\vec a \times \vec q\,|^2}(\vec{a}\times \vec q\,)_i(\vec{a}\times \vec q\,)_j\ .\nonumber
\end{align}
The condition $\eta^{\mu\nu}\Delta \tilde h_{\mu\nu}=0$ can be easily verified noticing that 
\be\label{RecurrenceRelation_Bessel}
J_0(x)+J_2(x)-2\frac{J_1(x)}{x}=0\ .
\ee

Plugging back in all the coefficients, from eqs. \eqref{htilde} and \eqref{Deltahtilde}  we obtain the exact expression for the FT of the metric in KS gauge, 
\be
\tilde{h}^{KN}_{\mu\nu}(\vec{q}\,) = -8\pi G M \tilde{h}_{\mu\nu}(\vec{q}\,) + {4\pi G Q^2} 
\Delta\tilde{h}_{\mu\nu}(\vec{q}\,) \ ,
\ee
which we will use in the rest of the paper to derive the amplitudes for particles scattering off a KN black hole. 

In the case of charged particles, we have to take into account the contribution to the amplitude coming from the electromagnetic interaction. To this end, we determine  the FT of the electric 4-potential of the KN solution,
\be
\hat{A}_\mu(q) =2 \pi \delta (q_0) \tilde{A}_\mu(\vec{q}\,) = 2 \pi \delta (q_0)\int d^3x e^{-i\vec{q}\cdot \vec{x}} {A}_\mu(x) \ ,
\ee
which, exploiting the same miracles as in the case of the gravitational field,  becomes
\be
\tilde{A}_\mu(\vec{q}\, ) = 4\pi Q \int_{q_\perp a}^\infty {udu \over {{|\vec{q}\,|}}^2} {K}_\mu(u, \vec{x}=i \partial_{\vec{q}}) j_0(u) \ .
\ee
The integrals are identical to the ones performed to compute $\tilde h_{0\mu}$ up to an overall factor, and the result is 
\be
\begin{aligned}
 & \tilde A_0(\vec q\,)=\frac{4 \pi Q}{{{|\vec{q}\,|}}^2}\cos(|\vec a \times \vec q\,|)\ ,\\  
 & \tilde{A}_i(\vec q\,)=-i\frac{4\pi Q}{{{|\vec{q}\,|}}^2}\left(\frac{q_i}{{{|\vec{q}\,|}}}\frac{\pi}{2}J_0(|\vec a \times \vec{q}\,|)-j_0(|\vec a \times \vec{q}\,|)(\vec{a}\times \vec q\,)_i\right)  \ . \label{Atilde}
\end{aligned}
\ee

\section{Tree-level Scattering Amplitudes and cross-sections}\label{Sec:Cross_Section}

The aim of this section is to use the exact expression for the metric in momentum space in KS gauge to derive formulae for tree-level scattering amplitudes for particles off a KN black hole. We will do this in detail for scalar probes, and we will briefly discuss the case of vector probes. Finally, for the special case of Kerr BHs we will derive the tree-level cross-section for the scattering of massive scalars. 

\subsection{Scalar probes}
\begin{figure}[h]
\centering
\includegraphics[width=0.30\textwidth, valign=c]{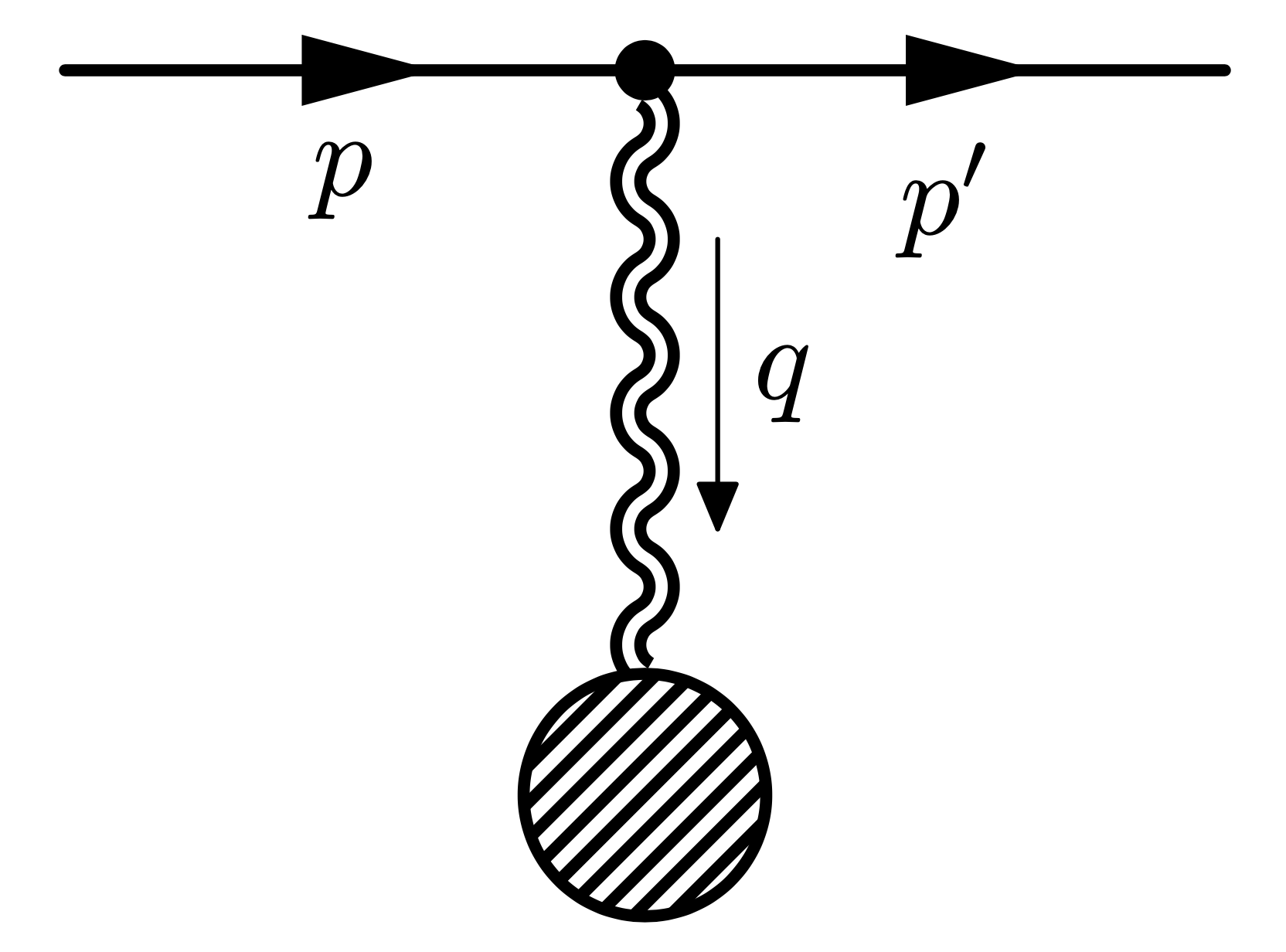}
\caption{Tree-level 1-graviton exchange diagram between the probe and the source.}
\label{Tree_Level_Amp}
\end{figure}

We start by considering a massive scalar field $\phi$ minimally coupled to the classical `on-shell' gravitational background  ${h}_{\mu\nu}(x)$. The remarkable property of the metric in KS gauge, namely $h= h_{\mu\nu} h^{\nu\lambda} =0$, implies that 
the `exact' coupling consists only of a tri-linear vertex 
\be
\mathcal{L}_{int}=\frac{1}{2}{h}^{\mu\nu}(x)T^\phi_{\mu\nu}(x)\ ,
\ee
where 
\be
T^\phi_{\mu \nu}(x)=\partial_\mu \phi \partial_\nu \phi - {1\over 2} \eta_{\mu\nu} 
(\partial\phi{\cdot}\partial \phi - m^2 \phi^2)
\ee
is the energy-momentum tensor of the scalar field. By performing the FT of ${h}_{\mu\nu}(x)$ to $\hat{h}^{KN}_{\mu\nu}(q)$ and contracting it with $T^{\phi}_{\mu\nu}$ in momentum space, which reads 
\be
\widetilde{T}^\phi_{\mu\nu}(p,p') = p_\mu p'_\nu + p_\nu p'_\mu - \eta_{\mu\nu}  (p{\cdot}p' - m^2)\ ,
\ee
we derive the sought for scattering amplitude 
\be
2\pi\delta (q_0)i\mathcal{M}_{KN}(p,p',\vec q\, )=\frac{i}{2}\hat{h}_{KN}^{\mu\nu}(q)\widetilde{T}^\phi_{\mu\nu}(p, p')=\frac{i}{2}2\pi\delta(q_0)\tilde{h}_{KN}^{\mu\nu}(\vec q\, )\widetilde{T}^\phi_{\mu\nu}(p, p') \label{amplitudeKerrnotsimplified}
\ee
depicted in Fig. \ref{Tree_Level_Amp}, where 
$q=p-p'$
is the transferred momentum and in general the external momenta are off-shell. Notice that the definition of the amplitude in eq. \eqref{amplitudeKerrnotsimplified} is a consequence of the fact that we are considering the scattering off a fixed stationary background. 
Using tracelessness of $\hat{h}_{\mu\nu}$ and noticing that $p_\mu=(E,-\vec p)$ one then finds
\be\label{amplitudeKerrsimplified}
i\mathcal{M}_{KN}(p,p',\vec q\, )=  i \tilde{h}_{KN}^{\mu\nu}(\vec q\, )  p_\mu p'_\nu = -i\Big(\tilde{h}^{KN}_{00} E E'+\tilde{h}^{KN}_{0i}(E'\vec p_i+E\vec p_i\!')+\tilde{h}^{KN}_{ij}\vec{p}_i \vec{p}_j\!'\Big)\ .
\ee

We can now give the amplitude in \eqref{amplitudeKerrsimplified} in a more explicit expression. Considering first the Kerr case, and observing that $\delta (q_0 )$ implies $E=E'$, from eq. \eqref{htilde} one gets
\be
\begin{aligned}
i \mathcal{M} (p, p', \vec q\, )&= i\frac{8\pi G M}{|\vec q\,|^2}\Bigg\{ E^2\cos|\vec{a}{\times}\vec{q}\,|+iE\Bigg(-\frac{\vec q \cdot (\vec p\,' + \vec p\,)}{|\vec q\,|}\frac{\pi}{2}J_0(|\vec{a}{\times}\vec{q}\,|) \\
&+ j_0(|\vec{a}{\times}\vec{q}\,|)(\vec{a}\times \vec{q}\,)\cdot( \vec p\, ' + \vec p\,)\Bigg) +{j_0(|\vec{a}{\times}\vec{q}\,|)}\left(\vec p \cdot \vec p\, '-2\frac{\vec q\cdot \vec p\ \vec q\cdot \vec p\, '}{|\vec q\,|^2}\right)\\
&-\frac{j_1(|\vec{a}{\times}\vec{q}\,|)}{|\vec{a}{\times}\vec{q}\,|}(\vec{a}\times\vec{q}\,)\cdot \vec p\ (\vec{a}\times\vec{q}\,)\cdot \vec p\, ' \\
&+\frac{1}{|\vec q\,|}\frac{\pi}{2}\frac{J_1(|\vec{a}{\times}\vec{q}\,|)}{|\vec{a}{\times}\vec{q}\,|}\Bigl(\vec q\cdot \vec p\, (\vec{a}\times \vec{q}\,)\cdot \vec p\, '+\vec q\cdot \vec p\, ' \ (\vec{a}{\times}\vec{q}\,)\cdot \vec p\Big)\Bigg\}\ , \label{Mamplitude}
\end{aligned}
\ee
where we notice that in general $p$ and $p'$ are off-shell. The presence of the terms proportional to the Bessel functions, which vanish on-shell, are a feature of the KS gauge and will play a crucial role when we will consider higher-loop amplitudes in section \ref{subleadingeikonalsection}.

It is straightforward to include also the contribution to the amplitude due to the electric charge. Using eq. \eqref{Deltahtilde} one finds
\begin{align}
i\Delta \mathcal{M}(p, p', \vec q\, )&=\left(-i \frac{4 \pi G Q^2}{{{|\vec{q}\,|}}}\right)\Bigg \{ E^2\frac{\pi}{2}J_0(|\vec a \times \vec q\,|)+iE\Bigg(-\frac{\vec q \cdot (\vec p + \vec p\, ')}{{{|\vec{q}\,|}}}j_0(|\vec a \times \vec q\,|)\nonumber  \\
&+\frac{\pi}{2}(\vec a \times \vec q\,)\cdot(\vec p+\vec p\, ')\frac{J_1(|\vec a \times \vec q\,|)}{|\vec a \times \vec q\,|}\Bigg)+\frac{\pi}{2} \frac{J_1(|\vec a \times \vec q\,|)}{|\vec a \times \vec q\,|}\Bigg(\vec p \cdot \vec p\,' -\frac{\vec q\cdot \vec p \ \vec q \cdot \vec p\,'}{{{|\vec{q}\,|}}^2}\Bigg)\nonumber \\
&+\frac{1}{{{|\vec{q}\,|}}}\frac{j_1(|\vec a \times \vec q\,|)}{|\vec a \times \vec q\,|}\Big(\vec q\cdot\vec  p\  (\vec a \times \vec q\,)\cdot \vec p\, '+\vec q\cdot\vec  p\, '\  (\vec a \times \vec q\,)\cdot\vec  p\Big) \label{DeltaMamplitude}\\
&-\frac{\pi}{2}\frac{J_2(|\vec a \times \vec q\,|)}{|\vec a \times \vec q\,|^2}(\vec a \times \vec q\,)\cdot \vec p\ (\vec a \times \vec q\,)\cdot \vec p\, '\Bigg \}\ ,\nonumber 
\end{align}
where now, with respect to eq. \eqref{Mamplitude}, the role of the functions $J_n$ and $j_n$ is exchanged, and in particular  the terms proportional to the spherical Bessel functions vanish on-shell. 

Finally, for a charged scalar we must also consider  the amplitude for the exchange of a photon in the KN background. 
Thanks to the KS condition $g_{\mu\nu}A^\mu A^\nu = \eta_{\mu\nu}A^\mu A^\nu=0$, there is again only a tri-linear interaction\footnote{Since $h^{\mu\nu}A_\nu=0$, one has $h^{\mu\nu}D_\mu\phi D_\nu\phi=h^{\mu\nu}\partial_\mu\phi \partial_\nu\phi$, which implies the gravitational coupling is unaltered by the interaction between the charged scalar and the electromagnetic field.} 
\be
{\cal L}_{int} = -j_\mu(x)A^\mu(x)=-i Q_\phi A^\mu (\phi^*\partial_\mu \phi - \phi\partial_\mu \phi^*)\ ,
\ee
where we have denoted with $Q_\phi$ the charge of the scalar.
This gives  rise to the amplitude
\be
i \mathcal {M}_A(p, p', \vec q\, )=-iQ_\phi \tilde{A}^\mu(\vec q\, ) (p_\mu +p'_\mu)=- iQ_\phi\Big((E+E')\tilde A_0 +\tilde A_i(\vec p _i+\vec p\, '_i)\Big)\ .
\ee
Using eq. \eqref{Atilde}, the amplitude reads
\be
\begin{aligned}
i\mathcal{M}_A(p, p', \vec q\, )&=\Bigg(-i\frac{4\pi Q Q_\phi}{{{|\vec{q}\,|}}^2}\Bigg)\Bigg\{2E \cos |\vec a \times \vec q\,|\\
&+i\Bigg(- \frac{\pi}{2}\frac{\vec q \cdot (\vec p + \vec p\, ')}{{{|\vec{q}\,|}}}J_0(|\vec a \times \vec q\,|)+j_0(|\vec a \times \vec q\,|)(\vec a \times \vec q\,)\cdot (\vec p+ \vec p\, ')\Bigg)\Bigg \}\ . \label{MAamplitude}
\end{aligned}
\ee

To summarise, the total amplitude for the tree-level scattering of a massive charged scalar off a KN black hole
reads 
\be
\mathcal{M}_{tot}=\mathcal{M}+\Delta \mathcal{M}+ \mathcal{M}_A\ ,\label{Mtot}
\ee
where $\mathcal{M}$, $\Delta \mathcal{M}$ and $\mathcal{M}_A$ are given in eqs. \eqref{Mamplitude}, \eqref{DeltaMamplitude} and \eqref{MAamplitude} respectively.

\subsection{Vector probes}

It is not difficult to generalize the analysis above to massive spinning probes. Consider in particular a spin $s=1$ (vector field) probe. In the simple case of a neutral massless `photon', the minimal coupling to the gravitational background reads
\be
h^{\mu\nu} T^A_{\mu\nu} = h^{\mu\nu} (F_{\mu\alpha} F_{\beta\nu}\eta^{\alpha \beta} + {1\over 4} \eta_{\mu\nu} F^2) =
h^{\mu\nu} F_{\mu\alpha} F^{\alpha}{}_{\nu}\ ,
\ee
where again we have made use of the properties of the metric in KS gauge. The relevant scattering amplitude is then given by 
\be
{i\over 2}\hat{h}_{KN}^{\mu\nu}(q) \widetilde{T}^A_{\mu\nu}(p, \varepsilon; p' \varepsilon') = 
-i\hat{h}_{KN}^{\mu\nu}(q) [p_\mu p'_\nu  \varepsilon{\cdot} \varepsilon' + \varepsilon_\mu \varepsilon'_\nu  p{\cdot}p'
- p_\mu \varepsilon'_\nu  \varepsilon{\cdot} p' - p'_\mu \varepsilon_\nu  \varepsilon'{\cdot} p]\ ,
\ee
where $\varepsilon$ and $\varepsilon'$ denote the polarizations of the photon, satisfying  
$p{\cdot}\varepsilon=0=p'{\cdot}\varepsilon'$. In the Coulomb gauge $\varepsilon(p) = (0, \vec{\varepsilon}\,)$ with $\vec{p}{\cdot}\vec{\varepsilon}=0$ and $\vec{\varepsilon}{\cdot}\vec{\varepsilon}\,^*=1$ so that only two independent polarizations survive. 


For massive vector fields the mass term drops from the gravitational coupling, exactly as in the scalar case, while in the case of a charged massive vector one has to consider the minimal coupling to the electromagnetic background, in which now a quartic coupling is present, contrary to the scalar case.


\subsection{Tree-level cross-section for scalars in a Kerr background}

In this subsection we write down the tree-level cross-section for the scattering of scalars, restricting our analysis to the Kerr case, since at this level the charged sector does not present any interesting difference. We define the 4-vector $\ell$ as the sum of the incoming and outgoing momenta, $\ell = p+p'\ ,$ where $q \cdot \ell = -\vec q\cdot \vec \ell = 0$ due to the condition $E=E'$. 
From eq. \eqref{Mamplitude}, imposing that the momenta are on-shell gives the on-shell tree-level amplitude in Fig. \ref{Tree_Level_Amp}, which reads
\begin{align}\label{Kerr_Tree_OnShell_Amp}
i\mathcal{M}_{on-shell}&= i\frac{8\pi G M}{{{|\vec{q}\,|^2}}}\Bigg\{ \cos |\vec a\times\vec q\,|\left(E^2+\frac{1}{4}\frac{(\vec a \times \vec q \cdot \vec \ell\,)^2}{|\vec a \times \vec q\,|^2}\right)\nonumber\\
&+\frac{\sin |\vec a \times \vec q\,|}{|\vec a\times \vec q\,|}\left({|\vec {p}\, |}^2-\frac{1}{4}\frac{(\vec a\times \vec q\cdot \vec \ell\,)^2}{|\vec a \times \vec q\,|^2}+iE\vec a\times \vec q\cdot \vec \ell  \right) \Bigg\}\ .
\end{align}
Notice that the Bessel functions disappear in the on-shell amplitude, while they contribute to higher-order terms. Instead, in the BH charge contribution to the amplitude in \eqref{DeltaMamplitude}, the roles between Bessel and spherical Bessel functions are inverted, and in the on-shell version of $\Delta\mathcal{M}$ the $J_n$'s are the ones that survive. 
Finally, the cross-section for scattering off a fixed (Kerr BH) target is given by
\be
d\sigma=\frac{1}{2 E v}2 \pi \delta(E-E')\left|\mathcal{M}_{on-shell}\right|^2\frac{d^3\vec p\, '}{(2 \pi)^32 E'}\ ,
\ee
with $v=|\vec p\,|/E$ the speed of the scattered particle and where the integration over the energy of the outgoing particle gives

\be
\frac{d\sigma}{d\Omega}=\frac{1}{16 \pi^2}|\mathcal{M}_{on-shell}|^2 \ .\label{treelevelcrosssection}
\ee

\begin{figure}[h]
\centering
\includegraphics[width=0.7\textwidth, valign=c]{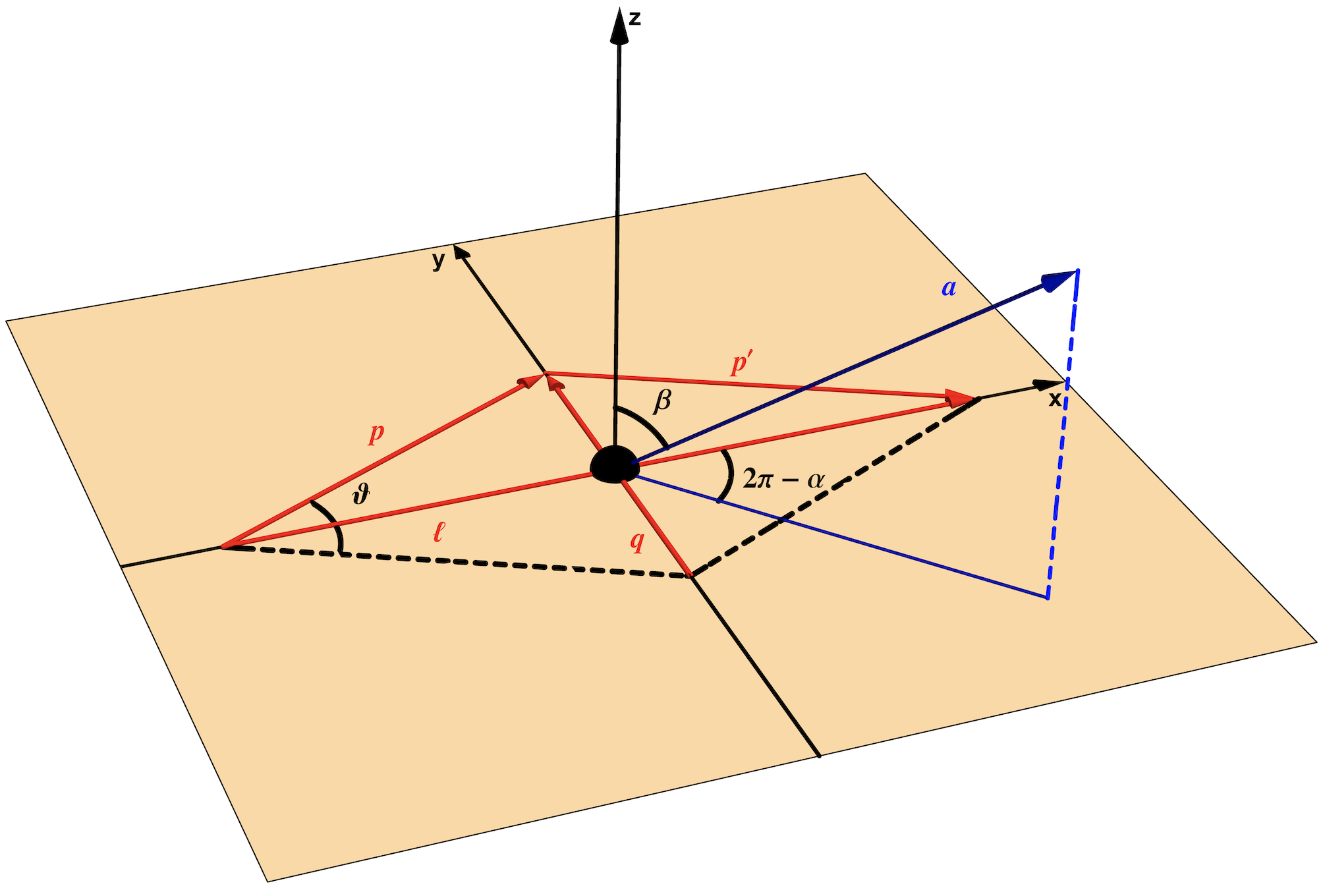}
\caption{Reference frame used to explicitly express the cross-section. }
\label{3D_Plane}
\end{figure}

We can now set a reference frame to explicitly express the cross-section in terms of angles. Instead of choosing the BH angular momentum along the $z$-axis,  we choose a reference frame in which the scattering occurs on the $x\text{-}y$ plane and the BH angular momentum has an arbitrary direction. In particular we choose the vector $\vec{q}$ to be along the $y$-axis and the vector $\vec \ell$ along the $x$-axis. However, it is important to notice that in general the geodesics in a Kerr space-time are non-planar, so neglecting the kinematics near the BH, we refer here and after to $\vec p$ and $\vec p\, '$ as asymptotic momenta, which define the scattering plane.
Therefore, denoting with $\vartheta$ the deflection angle, {\it i.e.} the angle between $\vec p$ and $\vec p\,'$, and with $a$ the modulus of $\vec{a}$, the kinematics reads
\be
\begin{gathered}\label{Refernce_Frame}
\vec{q}=2{|\vec{p}\,|}\sin\frac{\vartheta}{2}(0, 1, 0)\ , \qquad \vec{\ell}= 2{|\vec{p}\,|}\cos\frac{\vartheta}{2}(1, 0, 0)\ , \\
\vec{a}={{a}}\Big(\sin \beta \cos \alpha , \sin \beta \sin \alpha , \cos \beta \Big)  \ ,
\end{gathered}
\ee
as represented in Fig. \ref{3D_Plane}.
However, the cross-section must be considered only in the limit in which $|\vec q\, |\rightarrow 0$, which corresponds to the limit of small deflection angles $\vartheta\rightarrow 0$. In this regime then $\vec \ell\approx \vec p/2\approx\vec p\, '/2$, and the following relations are verified\footnote{Another point of view is that $\ell$, $\vec p$ and $\vec p\, '$ differ only by quantum corrections (see section \ref{Sec:Eik_Expansion}), which can be neglected since we are interested in classical quantities.}
\be
\begin{aligned}
|\vec a \times \vec q\,|^2&=4{{a}}^2{|\vec p\,|}^2 \sin^2\frac{\vartheta}{2}(\cos^2\beta+\cos^2\alpha\sin^2\beta)\\
&\approx a^2 {|\vec p\,|}^2 \vartheta^2(\cos^2\beta+\cos^2\alpha\sin^2\beta) \ ,
\end{aligned}
\ee
\be
 \vec a \times \vec q\cdot \vec \ell = -2{{a}}|\vec p\,|^2\sin\vartheta\cos \beta\approx -2{{a}}|\vec p\,|^2\vartheta\cos \beta\ .
\ee

Replacing these relations in \eqref{Kerr_Tree_OnShell_Amp} one gets 
\begin{align}
&{d\sigma\over d\Omega}\Big|_{Kerr} = \frac{G^2 M^2}{4v^4\sin^4(\vartheta/2)}
 \bigg\vert \cos |\vec a\times\vec q\,|\left(1+\frac{(\vec a \times \vec q \cdot \vec{v}\,)^2}{|\vec a \times \vec q\,|^2}\right)\nonumber\\
&+\frac{\sin |\vec a \times \vec q\,|}{|\vec a\times \vec q\,|}\left({{v}}^2-\frac{(\vec a\times \vec q\cdot \vec{v}\,)^2}{|\vec a \times \vec q\,|^2} +{2i}\vec{a}\times \vec{q}\cdot \vec{v}\,\right)  \bigg\vert^2\ ,\label{crosssectionkerr}
\end{align}
where we notice that by dimensional analysis, the $\vec{q}\,$'s in this expression should be thought of as wave vectors instead of momenta. Just by sanity check, in the non-rotating case (Schwarzschild BH) one can verify that the cross-section is nothing but the usual Rutherford-like formula
\be
\frac{d \sigma}{d\Omega}\Big|_{a=0}=\frac{G^2 M^2(1+v^2)^2}{4v^4\sin^4(\vartheta/2)} \ ,
\ee
that turns out to be independent of the energy of the probe and to scale with the area of the horizon.

To conclude, in eq. \eqref{crosssectionkerr} we have written the spherical Bessel functions in terms of trigonometric functions in order to make the comparison with the results of \cite{Bautista:2021wfy} more straightforward. In fact our cross-section and the one given in \cite{Bautista:2021wfy}  agree up to replacing $|\vec v||\vec a \times \vec q\, |\longleftrightarrow\vec a \times \vec q \cdot \vec v$. This does not mean that $\vec a \times \vec q \parallel \vec v$, but just that the expression in \cite{Bautista:2021wfy} must be considered in impact parameter space, where the transferred momentum is integrated and local terms can be dropped. This, as well as the extraction of the classical terms assigning powers of $\hbar$ to exchanged or loop momenta, will be discussed in the next section. 

\section{The eikonal expansion}\label{Sec:Eik_Expansion}

\begin{figure}[h]
\centering
\includegraphics[width=0.4\textwidth, valign=c]{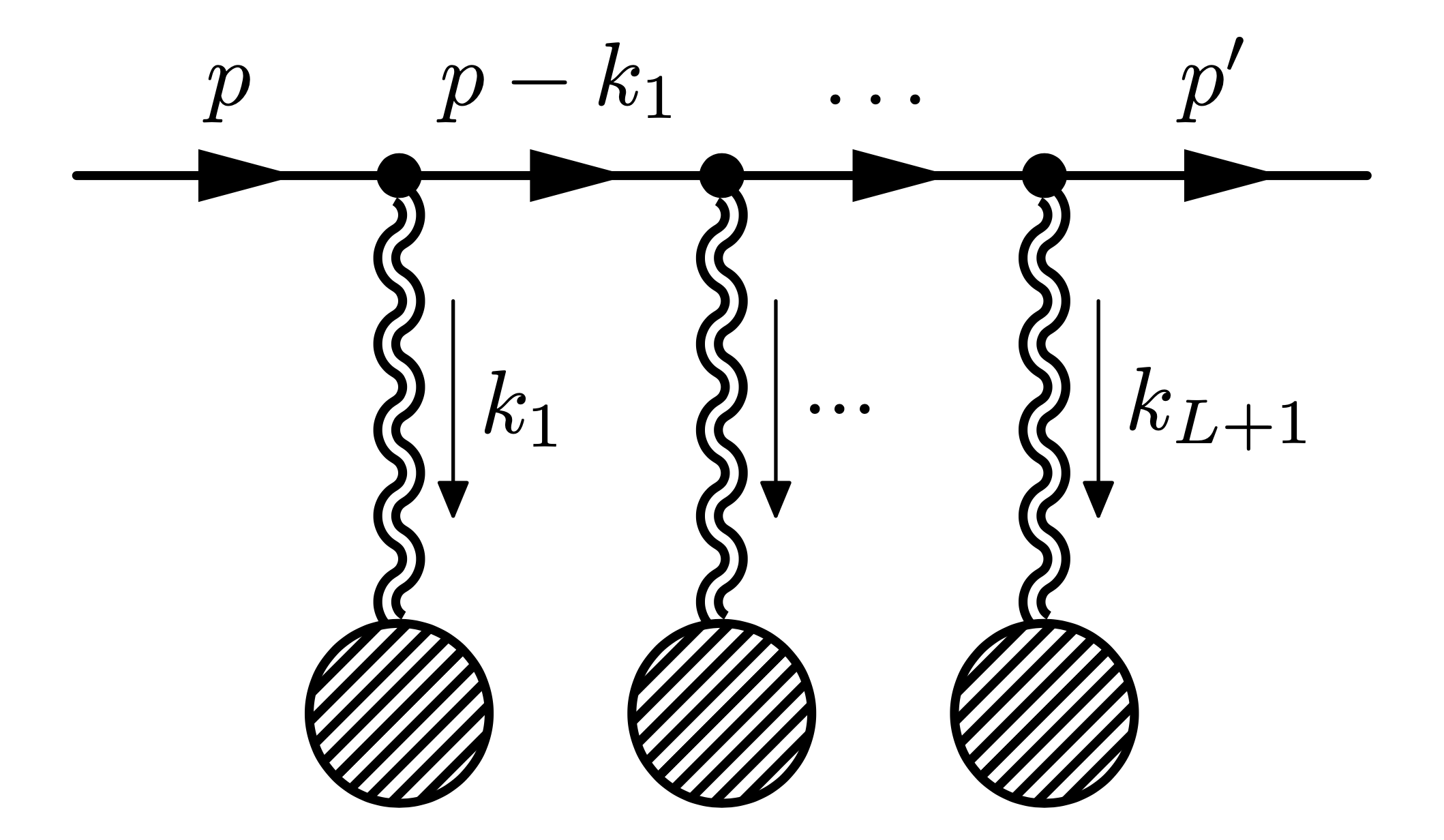}
\caption{Full scattering amplitude of $L+1$-gravitons exchange between the probe and the source.}
\label{L_loop_Amp}
\end{figure}

In this section we will show how the tree-level scattering amplitudes derived above allows us to determine  the leading contribution to the scattering angle, as well as sub-leading corrections, using the eikonal expansion. 
The crucial property of the KS gauge, namely the fact that the exact metric is first order in $G$, and therefore the only interaction vertex between the scalar and the metric is the one in Fig. \ref{Tree_Level_Amp}, has the remarkable consequence that at $L+1$-th order in the $G$ expansion the relevant diagrams in the classical limit are just comb-like diagrams with single tri-linear vertices inserted $L+1$ times on the probe world-line, as in Fig. \ref{L_loop_Amp} \cite{Menezes:2022tcs}. 
The corresponding $L$-loop scattering amplitude is obtained from the tree-level building block in \eqref{amplitudeKerrsimplified} by
\be\label{Amp_Lloop_generic}
i\mathcal{M}^{(L+1)}=\int \prod_{i=1}^L\frac{d^3 k_i}{(2\pi)^3}\prod_{i=1}^{L+1}i\mathcal{M}_{KN}(p_{i-1}, p_i,\vec{k}_i\, ) \prod_{i=1}^L\frac{i}{p_i^2-m^2+i\varepsilon} \ ,
\ee
where $p_{i-1} = p -\sum_{j=1}^{i-1}k_j$  with $p_i = p_{i-1} -k_i$ and $k_{L+1}=q-\sum_{i=1}^Lk_i$. Notice that in \eqref{Amp_Lloop_generic} the temporal components of the internal momenta are integrated out since each vertex carries a $\delta(k_i^0)$, as eq. \eqref{amplitudeKerrnotsimplified} shows.

As already mentioned, in order to connect the above scattering amplitudes to gravitational observables we consider the standard eikonal approach \cite{Amati:1987wq, Amati:1987uf, Amati:1990xe, Verlinde:1991iu, Kabat:1992tb, Levy:1969cr, AccettulliHuber:2020oou}. In the eikonal approximation we write the S-matrix as
\be
\widetilde{\cal S}(p,\vec b\,)=1+ i \widetilde{\cal T}(p,\vec b\,) = e^{2 i \delta (p,\vec b\,)}\ ,
\ee
where the eikonal phase $\delta (p, \vec{b}\, )$ is a function of $p$ and the impact parameter $\vec{b}$.
Expanding perturbatively in $G$ one gets
\be
i \widetilde{\cal T}(p, \vec b\,)=i \sum_{n=1}^{+\infty}\widetilde{\mathcal{M}}^{(n)}(p, \vec b\,)=\sum_{m=1}^{+\infty}\frac{1}{m!}\left(2i\sum_{n=1}^{+\infty}\delta^{(n)}(p, \vec b\,)\right)^m\ , \label{seriesexpeikonal}
\ee
where the index $n$ in both terms organizes the Post-Minkowskian (PM) expansion and   $i \widetilde{\mathcal{M}}^{(n)}$ is the amplitude in impact parameter space 
\be\label{Amp_in_IPS}
\widetilde{\mathcal{M}}^{(n)}(p, \vec{b}\,)=\frac{1}{2 {|\vec{p}}\,|}\int \frac{d^2 q}{(2 \pi)^2}e^{i\vec q \cdot \vec b} \mathcal{M}^{(n)}  \ ,
\ee
where we are integrating $\vec q$  on the plane orthogonal to the longitudinal momentum $\vec \ell$.

From the eikonal phase one can extract many physical observables, one of which is the deflection angle between the incoming and outgoing scattered particles, defined as  \cite{Amati:1990xe}
\be\label{Deflection_Ang}
\vartheta  (p, \vec b\, )=-\frac{2}{|\vec p\, |}\frac{\partial \delta (p, \vec b\, )}{\partial b}\ ,
\ee
where here and in the following we identify $b=|\vec b\, |$. At each loop order $L$, by computing the relevant scattering amplitude and using eq. \eqref{seriesexpeikonal} expanded up to $L+1$ PM order, one determines the eikonal phase and  therefore the scattering angle. At 1PM, the expansion of \eqref{seriesexpeikonal}  simply gives
\be
\widetilde{\mathcal{M}}^{(1)}(p, \vec{b}\,) = 2 \delta^{(1)}(p, \vec{b}\,) \ , \label{1PM_Eikonal_Phase}
\ee
while at 2PM one gets
\be\label{2PM_Eikonal_Formal}
\widetilde{\mathcal{M}}^{(2)}(p, \vec{b}\,)=2 \delta^{(2)}(p, \vec{b}\,)-\frac{i}{2}\left(2i\delta^{(1)}(p, \vec{b}\,)\right)^2\ ,
\ee
and similarly one obtains the relation between the amplitude and the eikonal phase at higher PM orders. 

One can observe that, starting from 2PM, the amplitude in impact parameter space leads to different contributions. The terms made by the combination of lower PM orders, which are called hyper-classical terms, come  from the fact that the eikonal phase is exponentiated. The classical terms instead are associated to the actual PM expansion of the eikonal phase and are the only non-trivial terms. 
In the usual way in which the amplitudes are computed in the literature using on-shell techniques, the classical terms arise from massive-particle irreducible amplitudes, while hyper-classical terms arise from amplitudes that are massive-particle reducible.
Therefore the classical term at $L+1$ PM order naturally contains the $L+3$-point amplitude with 2 massive scalars and $L+1$ gravitons, which embeds the knowledge of a contact vertex. On the other hand, in our KS-gauge approach none of the vertices with more than one graviton are present, and the full eikonal expansion must be derived from the diagrams in Fig. \ref{L_loop_Amp} using the amplitude \eqref{amplitudeKerrsimplified}.
Our goal here is to show how such computation is performed and organized.

At the end of the day the eikonal phase will contain both classical and quantum contributions, and we want to extract the classical one by following the KMOC formalism outlined in \cite{Kosower:2018adc}. Since the eikonal phase is related to the classical action via 
\be
\delta \sim \frac{1}{\hbar}S_{cl.}\ ,
\ee
it scales like $O(1/\hbar)$, and therefore in order to get the classical contribution one has to select from the amplitude the right dependence on $\hbar$. To this end we consider each `internal' momentum as `quantum' by performing the substitution
\be
q, k_i\rightarrow\hbar q, \hbar k_i\ . 
\ee
To be more clear the dependence on $\hbar$ of the momenta is
\be
p_i \rightarrow \frac{1}{2}\ell+\frac{1}{2}\hbar q  - \hbar\sum_{j=1}^{i}k_j  \ ,
\ee
where $\vec \ell$ encodes the information of the physical momentum and is the quantity that is kept fixed in the kinematical process. 
Besides, 
in order to take into account the fact that the $\vec{q}\,$'s in eq. \eqref{crosssectionkerr} are actually wave vectors, we formally also rescale $a$ by an inverse power of $\hbar$,
\be
a \rightarrow \frac{1}{\hbar} a \ .
\ee
Moreover, an inverse power of $\hbar$ for each vertex has to be considered. Finally, the eikonal phase that one obtains taking into account these $\hbar$ rules is a function of the longitudinal momentum  $\ell$, which at each order in the PM expansion can be substituted with either the incoming or outgoing momentum up to terms which vanish in the classical limit.  Observe that the rules translate in impact parameter space in having a full expansion in $a/b$ for each term in the $GM/b$ expansion. This is formally consistent, although one should recall that cosmic censorship requires $a\le \sqrt{G^2M^2-GQ^2}$, so that in principle $a/b<GM/b$ and higher powers of $a/b$ should be subdominant with respect to $GM/b$.

Let us now sketch how to organize the calculations in this framework. At tree-level,  the `in' and `out' momenta of the scalar are on-shell, and eq. \eqref{Amp_Lloop_generic} trivially reduces to $i\mathcal{M}^{(1)}=i\mathcal{M}_{KN}(p, p-q, \vec{q}\, )$. Then, from \eqref{1PM_Eikonal_Phase}, we see that this amplitude in impact parameter space is directly associated to $\delta^{(1)}$, which is of order $O(1/\hbar)$. From the aforementioned replacement rules we can see that to obtain such contribution the amplitude needs to be of order $O(1/\hbar^3)$, which exactly corresponds to taking $i\mathcal{M}^{(1)}=i\mathcal{M}_{KN}(p, p, \vec{q}\, )$, and substituting $\vec \ell$ with $2\vec p$ (or $2\vec p\, '$), since they only differ by quantum corrections. 

At one loop, the amplitude is 
\be\label{Amp_2loops}
i\mathcal{M}^{(2)}=\int \frac{d^3 k}{(2\pi)^3}i\mathcal{M}_{KN}(p, p -k,\vec{k}\, ) \frac{i}{(p-k)^2-m^2+i\varepsilon} i\mathcal{M}_{KN}(p-k, p-q,\vec{q}-\vec k\, ) \ ,
\ee
and from \eqref{2PM_Eikonal_Formal} we can see that this amplitude leads to two different contributions.  The hyper-classical term in \eqref{2PM_Eikonal_Formal} is of the order $O(1/\hbar^2)$, and is reconstructed from 
\be
i\mathcal{M}^{(2)}\Big|_{hyp.\ cl.}=O(1/\hbar^4)\rightarrow i\widetilde{\mathcal{M}}^{(2)}\Big|_{hyp.\ cl.}=2 i\left(\delta^{(1)}\right)^2=O(1/\hbar^2)\ .
\ee
It can be proved that at each order, such hyper-classical terms are recovered by a convolution of the amplitudes in momentum space. For instance this can be easily seen for the 2PM hyper-classical term, where the relevant contribution comes from
\be
i \mathcal{M}^{(2)}\Big|_{hyp.\ cl.}=\frac{1}{2}\int\frac{d^3k}{(2\pi)^3}i\mathcal{M}_{KN}(p, p, \vec k\, )i \mathcal{M}_{KN}(p, p, \vec q-\vec k\, )2\pi \delta(\vec \ell\cdot \vec k\, )\ ,
\ee
where the $\delta(\vec \ell\cdot \vec k\, )$ is associated to the expansion of the propagator at order $O(1/\hbar)$ (see eq. 5.4 of \cite{Brandhuber:2021eyq}).
Now integrating over the transverse loop momenta and considering the amplitude in impact parameter space, one gets
\be
\begin{aligned}
i\widetilde{\mathcal{M}}^{(2)}\Big|_{hyp.\ cl.}&=\frac{1}{2}\frac{1}{4 p^2}\int\frac{d^2 q}{(2\pi)^2}\int \frac{d^2k}{(2 \pi)^2}e^{i\vec q\cdot \vec b}i\mathcal{M}_{KN}(p, p, \vec k\, )i\mathcal{M}_{KN}(p,p,  \vec q-\vec k\, )\\
&=\frac{1}{2}\left(i \widetilde{\mathcal{M}}^{(1)}\right)^2=\frac{1}{2}\left(2 i \delta^{(1)}\right)^2\ ,
\end{aligned}
\ee
exactly as expected in \eqref{2PM_Eikonal_Formal}, and for higher orders terms the derivation is very similar, even for mixed terms. 

The only non-trivial contribution to the eikonal phase are the classical contributions, and from the $\hbar$ expansion, at 2PM such term is associated to\footnote{This scaling of the classical contribution is actually completely general and holds at every loop order.}
\be
i \mathcal{M}^{(2)}\Big|_{cl.}=O(1/\hbar^3)\ .
\ee
At this point it is convenient to decompose the building block amplitude as follows
\be\label{M0_Mextra}
i\mathcal{M}^{KN}=i\mathcal{M}^{KN}_0+i\mathcal{M}^{KN}_{extra}\ ,
\ee
where $\mathcal{M}^{KN}_0$ contains all the terms of the amplitude that are non-vanishing on-shell, while in $\mathcal{M}^{KN}_{extra}$ there are additional terms, that vanish when the scalar momenta are on-shell, but contribute to \eqref{Amp_2loops}. Then the 1-loop amplitude will be decomposed in three different pieces, {\it viz.}
\begin{equation}\label{MKN_0extra}
    i\mathcal{M}^{(2)}\Big|_{cl.}=i\mathcal{M}^{(2)}\Big|_{cl.}^{(0, 0)}+i\mathcal{M}^{(2)}\Big|_{cl.}^{(extra, extra)}+2i\mathcal{M}^{(2)}\Big|_{cl.}^{(extra, 0)}\ .
\end{equation}
For the $(extra, extra)$ term, one can show that the contribution of the propagator is factored out, so that the amplitude looks like arising from a contact vertex, which we can represent schematically as
\be\label{Eq:ContactTerm}
i\mathcal{M}^{(2)}\Big|_{cl.}^{(extra, extra)} \sim \includegraphics[width=0.2\textwidth, valign=c] {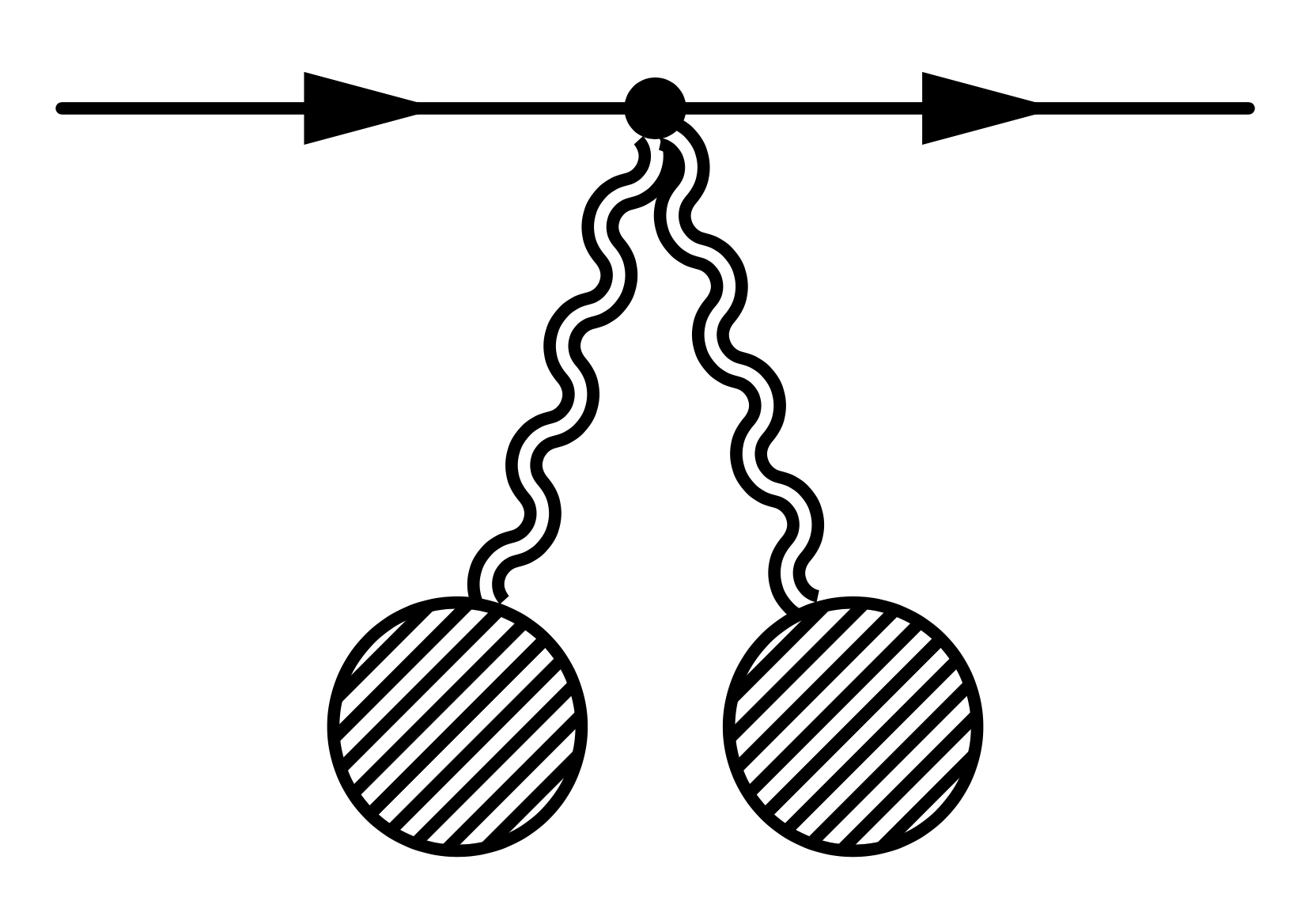} \ .
\ee
Remarkably, the mixed term gives an imaginary contribution which vanishes after integration. This is expected since we know that imaginary contributions are associated to radiative corrections that start at 3PM \cite{Damour:2020tta}.
Then the classical term turns out to be given by the $(0, 0)$ term, in which the propagator is not factored out, and the $(extra, extra)$ term, which is mimicking the contact vertices.

Restricting  to the case of Kerr BHs, the $\mathcal{M}_{0}$ and $\mathcal{M}_{extra}$ parts of the amplitude are respectively
\begin{align}
&i \mathcal{M}_0 (p_{i-1}, p_i, \vec k_i  )= i\frac{8\pi G M}{|\vec k_i|^2}\Bigg\{ E^2\cos|\vec{a}{\times}\vec{k}_i|+iE
 j_0(|\vec{a}{\times}\vec{k}_i|)(\vec{a}\times \vec{k}_i)\cdot( \vec p_i + \vec p_{i-1}) \label{Mamplitude0} \\
 & \ +{j_0(|\vec{a}{\times}\vec{k}_i|)}\left(\vec p_i \cdot \vec p_{i-1}-2\frac{\vec k_i\cdot \vec p_i\ \vec k_i\cdot \vec p_{i-1}}{|\vec k_i|^2}\right)
-\frac{j_1(|\vec{a}{\times}\vec{k}_i|)}{|\vec{a}{\times}\vec{k}_i|}(\vec{a}\times\vec{k}_i)\cdot \vec p_i (\vec{a}\times\vec{k}_i)\cdot \vec p_{i-1}\Bigg\} \ ,\nonumber 
\end{align}
and
\begin{align}
i \mathcal{M}_{extra} (p_{i-1}, p_i, \vec k_i  )&= i\frac{4\pi^2 G M}{|\vec k_i|^2}\Bigg\{ -iE\frac{\vec k_i \cdot (\vec p_i + \vec p_{i-1})}{|\vec k_i|}J_0(|\vec{a}{\times}\vec{k}_i|) \label{Mamplitudeextra} \\
&+\frac{1}{|\vec k_i|}\frac{J_1(|\vec{a}{\times}\vec{k}_i|)}{|\vec{a}{\times}\vec{k}_i|}\Bigl(\vec q\cdot \vec p_{i-1} (\vec{a}\times \vec{k}_i)\cdot \vec p_i+\vec k_i\cdot \vec p_i \ (\vec{a}{\times}\vec{k}_i)\cdot \vec p_{i-1}\Big)\Bigg\}\ , \nonumber
\end{align}
which in particular shows that the terms $\mathcal{M}_{extra}$ are those containing the Bessel functions $J_n$. One can analogously identify the same structure in the terms proportional to $Q^2$ for the KN BHs, that are given in eq. \eqref{DeltaMamplitude}. In this case one can show that the terms  $\Delta \mathcal{M}_0$ are instead the ones containing the Bessel functions $J_n$, while the terms $\Delta \mathcal{M}_{extra}$ contain spherical Bessel, {\it i.e.} trigonometric functions. 

In the next two sections we will show how this works explicitly. In particular in the next section we will consider the leading eikonal contribution for KN BHs, while in the following section we will analyze the first subleading correction, {\it i.e.} one loop, focusing in detail on the case of a Schwarzschild BH. Anyway, we expect that the construction be completely general and allow in principle to determine the deflection angle at any order in the PM expansion.

\section{Leading eikonal contribution}\label{Sec:Leading_Eik_Kerr}

In this section we will discuss how to derive the leading eikonal phase for probe scattering off KN BHs, taking into account also the effects due to the charge of the BH, and we will use it to evaluate the deflection angle. In the first part we review the Kerr case and compute the eikonal phase performing the $2d$-integral in the transferred momentum exactly, after having  neglected local terms. As a result we write down and analyze an exact expression (for every orientation) for the deflection angle.  
In the KN case the integration is more subtle, and a more suitable alternative approach is proposed. In particular, introducing an auxiliary integration variable, we compute the eikonal phase by performing a $3d$-integral in the transferred momentum exploiting the properties of the KS gauge.
We then determine the deflection angle as an expansion in orders of the inverse impact parameter, and we comment on  the results. Finally we discuss the effect of the gauge potential in the case in which the probe is charged, and we argue how for reasonable energies such contribution will be dominant in the dynamics.   

\subsection{Kerr case}
\begin{figure}[h]
\centering
\includegraphics[width=0.7\textwidth, valign=c]{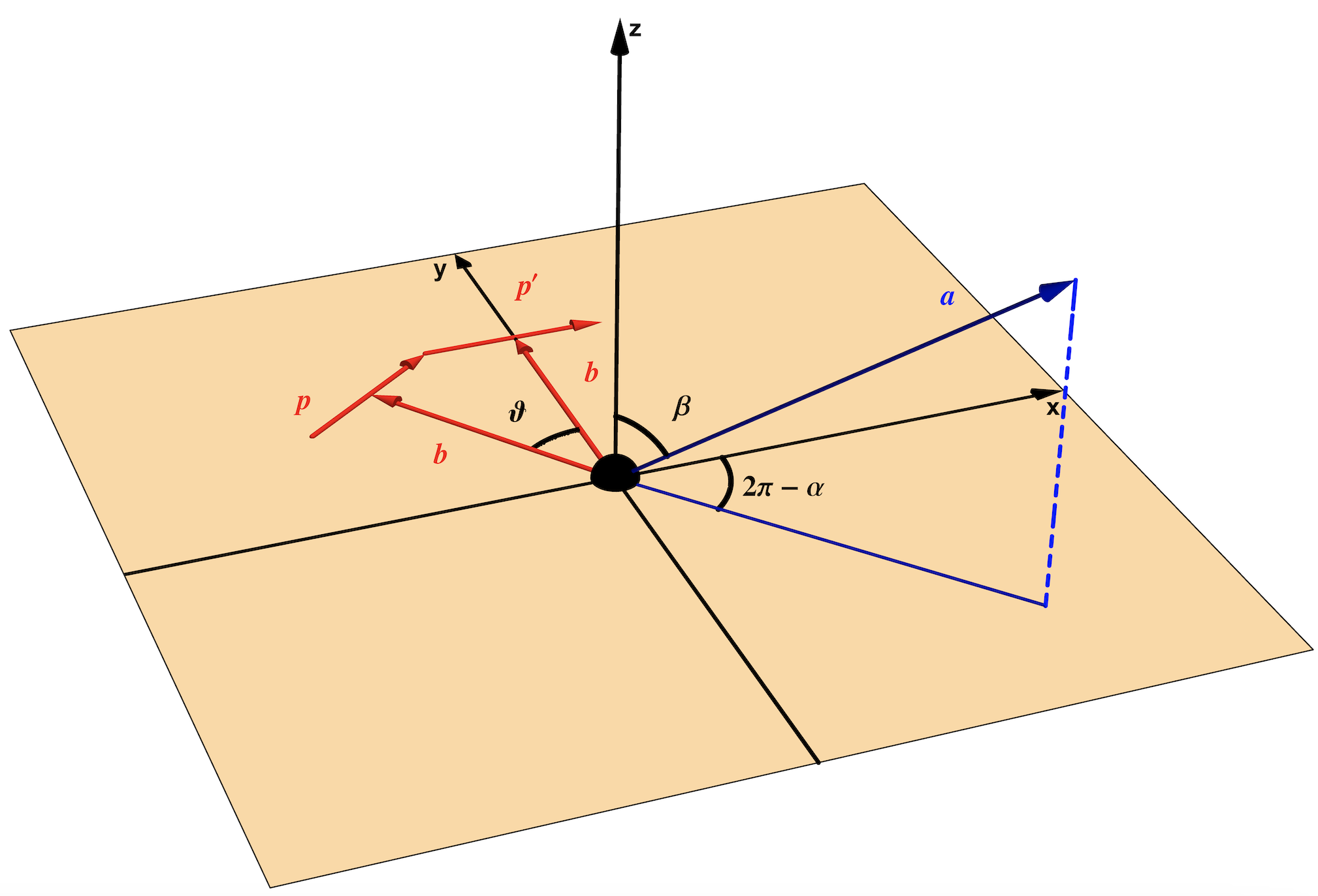}
\caption{Reference frame used to explicitly express the deflection angle.}
\label{3D_Plane_Eikonal}
\end{figure}

\begin{figure}[h]
\centering
\includegraphics[width=0.48\textwidth, valign=c]{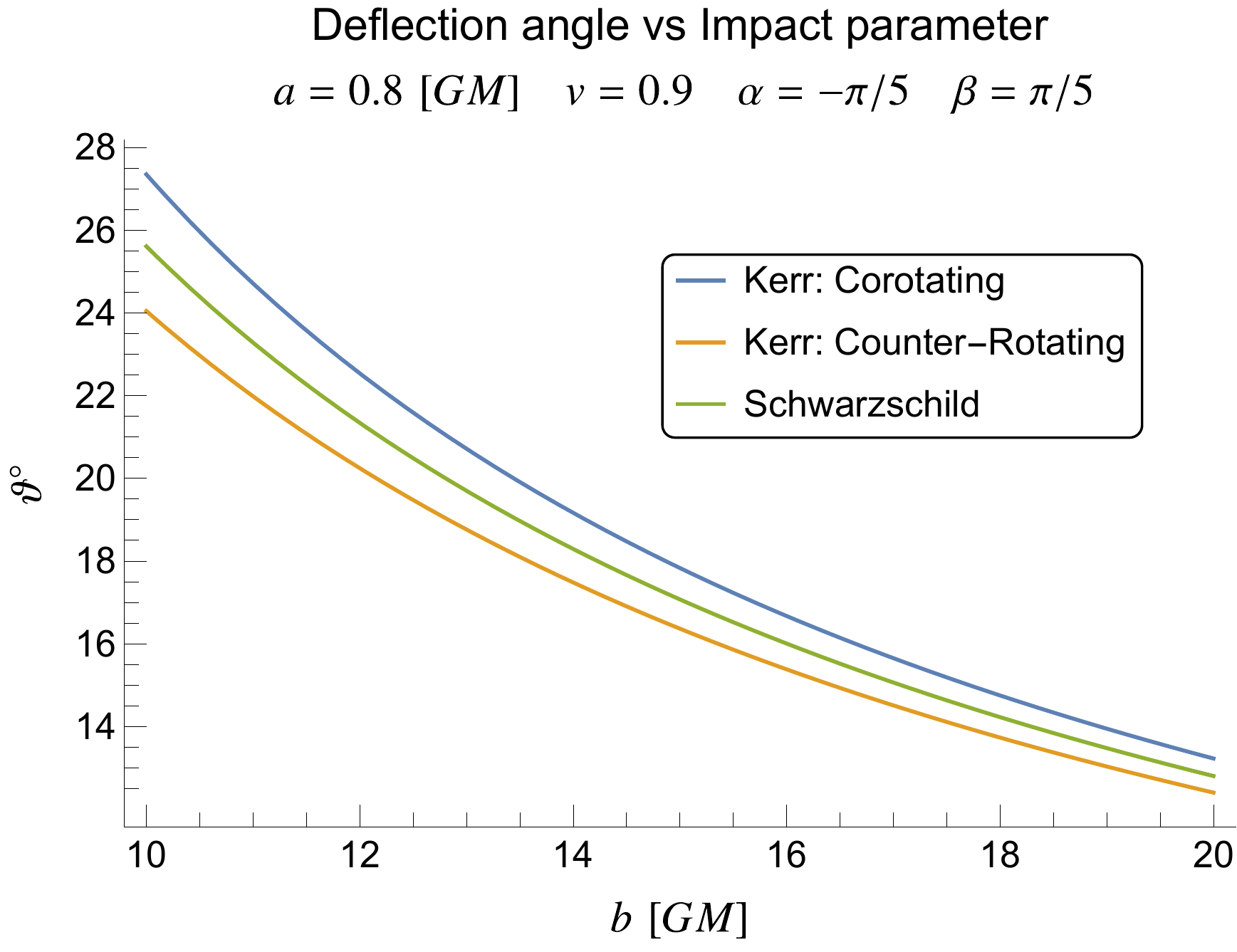}
\includegraphics[width=0.48\textwidth, valign=c]{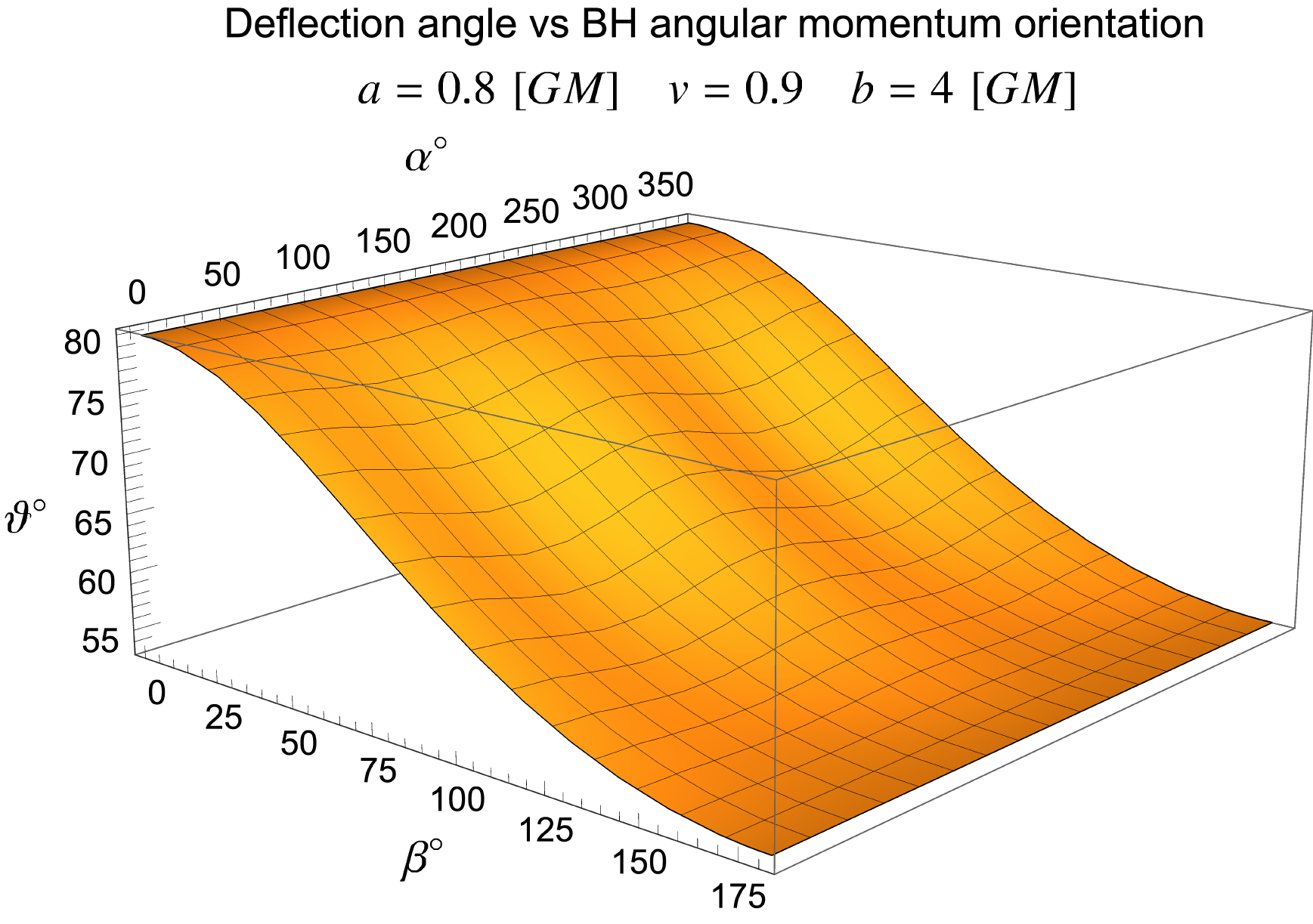}
\caption{Study of the deflection angle at 1PM for Kerr BHs. On the left the comparison of the deflection angle vs the impact parameter, with $v$ and $\vec a$ fixed. On the right the deflection angle vs the orientation of the BH angular momentum for fixed $v$, $b$ and $a$.}
\label{DeflectionAng_Kerr_Plot}
\end{figure}

Considering the eikonal leading order ($L=0$) means taking the on-shell version of \eqref{Mamplitude}, which exactly corresponds to $\mathcal{M}_{on-shell}$  in eq. \eqref{Kerr_Tree_OnShell_Amp} switching to impact parameter space with the help of \eqref{Amp_in_IPS}. Since we are interested in the long-range regime, we can neglect local terms, which arise from terms $O(|\vec q\, |^2)$ in the numerator of the amplitude when integrating over the transferred momentum. This means that under integration we can perform the replacements
\be
|\vec a \times \vec q\,|^2=a^2 |\vec q\, |^2-(\vec q \cdot \vec a)^2\rightarrow -(\vec q \cdot \vec a)^2\ 
\ee
and
\be
|\vec q\times (\vec \ell \times \vec a)|^2=|\vec q\, |^2 |\vec \ell \times \vec a\,|^2-(\vec q\cdot \vec \ell \times \vec a)^2\rightarrow -(\vec q\cdot \vec \ell \times \vec a)^2\ .
\ee
Moreover, since $\vec \ell \perp \vec q$, one has
\be
|\vec q\times (\vec \ell \times \vec a)|^2=|\vec \ell\, |^2(\vec q \cdot \vec a)^2\ ,
\ee
implying that we can perform the additional replacement
\be\label{LocalTerms_Simplification}
|\vec a \times \vec q|\rightarrow\pm\, \vec q\cdot \hat \ell \times \vec a\ . 
\ee
Inserting \eqref{LocalTerms_Simplification} into \eqref{Kerr_Tree_OnShell_Amp} the ambiguity of the sign in eq. \eqref{LocalTerms_Simplification} disappears and the leading eikonal phase reads 
\be\label{Leading_Eikonal_Kerr_Implicit}
\delta^{(1)}(p, \vec b)=\frac{8\pi G M}{4|\vec p\, |}\int\frac{d^2q}{(2\pi)^2}\frac{e^{i\vec q\cdot \vec b}}{{{|\vec{q}\,|^2}}}\Bigg\{ \cos( \vec q\cdot \hat \ell \times \vec a)\left(E^2+|\vec p\, |^2\right)+2iE|\vec p\, |\sin( \vec q\cdot \hat \ell \times \vec a) \Bigg\}\ ,
\ee
where we have replaced $\vec \ell\rightarrow \vec p$, which differ only by quantum corrections. 
In order to perform the FT, one can make use of the master integral
\be
\label{IPS_Integral}
\mathcal{F}(d, \nu) =\int\frac{d^dq}{(2 \pi)^d}\, e^{i\vec q \cdot \vec x} {{|\vec{q}\,|}}^{2\nu} = \frac{2^{2\nu}}{\pi^{d/2}}\frac{\Gamma(\nu+d/2)}{\Gamma(-\nu)}\frac{1}{{{|\vec{x}\,|}}^{2\nu+d}}\  ,
\ee
valid for $\nu \neq -d/2$, which for $\nu = -d/2$ needs to be regulated by the introduction of an energy cut-off  $\mu$ so that one gets 
\be \label{IPS_Integrallog}
\mathcal{F}(d, -d/2) =\int\frac{d^dq}{(2 \pi)^d}\, e^{i\vec q \cdot \vec x} {{|\vec{q}\,|}}^{-d} = -\frac{2^{1-d}}{\pi^{d/2} \Gamma(d/2)} \log \mu |\vec{x}\,| \ .
\ee
In particular for $d=2$ and $\nu=-1$  one finds
\be
\mathcal{F}(2, -1)=-\frac{1}{2\pi}\log \mu |\vec x\, |\ .
\ee
Finally, by writing the trigonometric functions in exponential form one gets
\be\label{Cos_IPS}
\int\frac{d^2q}{(2\pi)^2}\frac{e^{i\vec q\cdot \vec b}}{|\vec q\, |^2}\cos(\vec q\cdot \hat \ell \times \vec a)=-\frac{1}{4\pi}\log \mu^2|\vec b-\hat \ell \times \vec a||\vec b +\hat \ell \times \vec a|\ ,
\ee
\be\label{Sin_IPS}
\int\frac{d^2q}{(2\pi)^2}\frac{e^{i\vec q\cdot \vec b}}{|\vec q\, |^2}\sin(\vec q\cdot \hat \ell \times \vec a)=-\frac{1}{4\pi\, i}\log\frac{|\vec b+\hat \ell \times \vec a|}{|\vec b -\hat \ell \times \vec a|}\ .
\ee
Plugging these results in eq. \eqref{Leading_Eikonal_Kerr_Implicit},  we can write the eikonal phase  in a very compact form  as
\be\label{delta_in_LogSum}
\delta^{(1)}(p, \vec b)=-\frac{G M E^2}{2 |\vec p\, |}\sum_{\pm}(1\pm v)^2\log \mu|\vec b\pm \hat p\times \vec a|\ ,
\ee
which exactly corresponds to the result in \cite{Guevara:2018wpp}. We can finally consider the deflection angle at 1PM using \eqref{Deflection_Ang}. With an explicit parametrization of the vectors like in Fig. \ref{3D_Plane_Eikonal} in which
\be
|\vec b\pm \hat p\times \vec a|^2= a^2\sin^2\alpha\sin^2\beta+(b\mp a\cos\beta)^2\ ,
\ee
the result, plotted in Fig. \ref{DeflectionAng_Kerr_Plot},  reads
\be\label{theta_Kerrcase}
\vartheta^{(1)}=\frac{GM}{v^2}\sum_{\pm}\frac{(1\pm v)^2(b\mp a \cos\beta )}{a^2\sin^2\alpha \sin^2 \beta+(b\mp a \cos \beta)^2}\ .
\ee
It is important to notice that the above expression depends only on the velocity of the probe, while one would expect a behavior like $\vartheta\sim 1/(E b)$. This is due to the fact that since we are considering a gravitational process, the dimensionless expansion parameter is $GME$, which exactly cancels the energy dependence that would appear in eq. \eqref{theta_Kerrcase}.

We can now briefly analyse the plots in Fig. \ref{DeflectionAng_Kerr_Plot}.
From the left panel of the figure,  we can appreciate the difference between the co-rotating and counter-rotating case, with the deflection angle being larger  in the former case as expected. In the right panel this behaviour can be seen more clearly, in addition to the fact that the deflection angle is more sensitive to the polar angle $\alpha$ when $\beta=\pi/2$, namely when the BH axis lies on the scattering plane. 

It is worth noticing that we can express the deflection angle with respect to either $\vec p$ or $\vec p\, '$. In fact, while in the former case we assume to know the direction of $\vec p$ with respect to the orientation of the BH, in the latter  we imagine to see the direction of $\vec p\, '$ which is deflected from its original trajectory, which is a more natural scenario from an experimental point of view. In the end, in this probe approximation, the BH angular momentum remains unchanged, and therefore  the angular momentum of the probe is conserved, which means that  $\vec p$, $\vec p\, '$ and the BH lie on the same plane,
so that the deflection angle $\vartheta$ fully reconstructs the scattering process. 
Finally, considering the limit case in which $a=0$
\be
\vartheta^{(1)}\Big|_{a=0}=\frac{2GM}{v^2\, b}(1+v^2) \ ,
\ee
and also $m=0$
\be
\vartheta^{(1)}\Big|_{\substack{m=0 \\ a=0}}=\frac{4GM}{b}\ ,
\ee
we recover the celebrated Einstein's deflection formula.

To conclude, the procedure followed to obtain eq. \eqref{delta_in_LogSum}, namely making use of the replacement \eqref{LocalTerms_Simplification}, is not general but it holds only for the Kerr case. In fact the key point in the discussion was that in $\mathcal{M}_{on-shell}$ there is an overall dependence of $1/|\vec q\, |^2$, which allows to make the replacement \eqref{LocalTerms_Simplification} up to terms $|\vec q\, |^{2 n}$. Such terms, as eq. \eqref{IPS_Integral} shows, are vanishing\footnote{Actually they are local terms, namely delta-functions and their derivatives, which we neglect since we are interested only in the long-range regime.}, and make the aforementioned substitution legal under the integration over the transferred momentum. For what concerns the KN case, at first glance we would be tempted to use eq. \eqref{LocalTerms_Simplification} also on the on-shell version of \eqref{Deltahtilde}, which reads
\begin{equation}\label{ChargeAmp_OnShell}
\begin{aligned}
i\Delta\mathcal{M}_{on-shell}(p, \vec q\, )&=-i\frac{4 \pi G Q^2}{|\vec q\, |}\Bigg\{E^2J_0(|\vec a \times \vec q\, |)-\frac{J_2(|\vec a \times\vec q\, |)}{|\vec a \times \vec q\, |^2}(\vec a\times \vec q\cdot \vec p\, )^2\\
&+\frac{J_1(|\vec a \times\vec q\, |)}{|\vec a\times \vec q\,|}\Big(|\vec p\, |^2+2iE\vec a\times\vec q\cdot \vec p\Big)\Bigg\} \ ,
\end{aligned}
\end{equation}
and simplify the amplitude. However, the overall factor $1/|\vec q\, |$ invalidates the replacement in this case. The reason is that expanding the Bessel functions in eq. \eqref{ChargeAmp_OnShell}, the amplitude will generate only even powers of the transferred momentum (like in the Kerr case), however the overall $1/|\vec q\, |$ produces odd powers at every order in the expansion. As we can see from \eqref{IPS_Integral}, the integral of odd powers of the transferred momentum does not vanish, which means that we cannot neglect any terms. For this reason we have to find an alternative way to express the eikonal phase in the KN case, which is discussed in the next subsection. 

\subsection{Kerr-Newman case}

We now discuss a different method to derive the eikonal phase at leading order. As already discussed in section \ref{Sec:Eik_Expansion}, by definition the eikonal phase at 1PM is given by
\begin{equation}
    \delta^{(1)}_{KN}(p, \vec b\, )=\frac{1}{4|\vec p\, |}\int\frac{d^2q}{(2\pi)^2}e^{i\vec q\cdot \vec b}\mathcal{M}_{on-shell}^{KN}(p, \vec q\, )\ ,
\end{equation}
where the orthogonality relations $\vec \ell \cdot \vec b=\vec \ell \cdot \vec q=0$ are satisfied, and we recall that up to quantum corrections $\ell\approx 2\vec p\approx2 \vec p\, '$. 
It is possible to express the eikonal phase in terms of a $3d$-FT thanks to 
\begin{equation}\label{EikPhase_3d}
    \delta^{(1)}_{KN}(p, \vec b\, )=\frac{1}{2}\int\frac{d^3q}{(2\pi)^3}e^{i\vec q\cdot \vec b}\, 2\pi\delta(\vec q\cdot \vec \ell\, )\mathcal{M}_{on-shell}^{KN}(p, \vec q\, )\ ,
\end{equation}
in which now $\vec q\cdot \vec \ell\neq 0$. Finally we can rewrite \eqref{EikPhase_3d} as a shifted FT \cite{Levy:1969cr} such that
\begin{equation}\label{EikPhase_3dFT}
\begin{aligned}
    \delta^{(1)}_{KN}(p, \vec b\, )&=\frac{1}{2}\int_{-\infty}^{+\infty}d\xi\int\frac{d^3q}{(2\pi)^3}e^{i\vec q\cdot (\vec b+\xi \vec \ell\, )}\mathcal{M}_{on-shell}^{KN}(p, \vec q\, )\\
&=\frac{1}{2}\int_{-\infty}^{+\infty}d\xi \ \text{FT}\Big[\mathcal{M}_{on-shell}^{KN}\Big](p, \vec b+\xi \vec \ell\, ) ,
\end{aligned}
\end{equation}
where $\xi$ is a real integration variable with dimension of squared length\footnote{Notice that eq. \eqref{EikPhase_3dFT} can be generalized to higher PM orders by replacing $\mathcal{M}_{on-shell}^{KN}$ with the classical contribution of the amplitude at the relevant loop order.}. 

We can explicitly consider \eqref{amplitudeKerrsimplified} that leads to 
\begin{equation}
    \text{FT}\Big[\mathcal{M}^{KN}_{on-shell}\Big](p, \vec b+\xi \vec \ell\, )=-h_{\mu\nu}(\vec x=\vec b+\xi \vec \ell\, )p^\mu p^\nu\ ,
\end{equation}
from which we can finally write
\begin{equation}\label{Eik_hFT_NotExplicit}
    \delta^{(1)}_{KN}(p, \vec b\, )=\frac{1}{2}\int_{-\infty}^{+\infty}d\xi\frac{2 G M r^3-GQ^2r^2}{r^4+z^2a^2}\left(K_{\mu}p^{\mu}\right)^2\Bigg|_{\vec x=\vec b+\xi \vec \ell}\ ,
\end{equation}
where we have used the original definitions in section \ref{Sec:FT}. Notice that eq. \eqref{Eik_hFT_NotExplicit} is written in OS coordinates, in which $r^2\neq x^2+y^2+z^2$, in fact following the shift replacement we can define 
\begin{equation}
    \vec B = \vec b+2\xi \vec p\ ,
\end{equation}
such that $|\vec B\, |^2=b^2+4\xi^2|\vec p\, |^2$, and
\begin{equation}
    \mathcal{B}^2=r^2\Big|_{\vec x=\vec B}=\frac{1}{2}\left(|\vec B\, |^2-a^2\right)+\frac{1}{2}\sqrt{\left(|\vec B\, |^2-a^2\right)^2+4\left(\vec a\cdot \vec B\, \right)} \ , 
\end{equation}
which makes it possible to rewrite the eikonal phase as
\begin{equation}\label{Eik_hFT_Explicit}
    \delta^{(1)}_{KN}(p, \vec b\, )=\frac{1}{2}\displaystyle\int_{-\infty}^{+\infty}d\xi\frac{2 G M \mathcal{B}^3-GQ^2\mathcal{B}^2}{\mathcal{B}^4+(\vec a \cdot \vec B\, )^2}\left(E+\frac{2\xi |\vec p\, |^2+\vec a \cdot \vec p\times \vec b}{\mathcal{B}^2+a^2}+\frac{\vec a \cdot \vec B\ \vec a \cdot \vec p}{\mathcal{B}\left(\mathcal{B}^2+a^2\right)}\right)^2 \ , 
\end{equation}
valid for an arbitrary BH angular momentum orientation. 

In the equation above, $\delta^{(1)}_{KN}$ is the sum of the two contributions 
\begin{equation}
   \delta^{(1)}_{KN}=\delta^{(1)}+\Delta\delta^{(1)} \ ,
\end{equation}
 where the first is proportional to the mass and was computed in the previous section, while  the second is the term proportional to the square of the charge of the BH. Evaluating the integral in \eqref{Eik_hFT_Explicit} would thus give a result that coincides with 
\eqref{delta_in_LogSum} when $Q=0$. 
However, the integral in eq. \eqref{Eik_hFT_Explicit} is quite involved,  and it is very difficult to compare it with the exact result in \eqref{delta_in_LogSum} for what concerns the Kerr case. Instead, we have checked the validity of \eqref{Eik_hFT_Explicit} by
computing it order by order in an expansion in $a$ (which is equivalent to an expansion in $1/b$),  and found perfect agreement up to very high order in the expansion. Moreover, it is amusing to observe that the exact result in \eqref{delta_in_LogSum} can be reconstructed from the equatorial limit, \textit{i.e.} the case in which $\vec a$ is along the $z$-axis referring to eq. \eqref{Refernce_Frame}. In fact in this case eq. \eqref{Eik_hFT_Explicit} can be easily integrated, and for the Kerr case one obtains
\be
\delta^{(1)}(p, \vec b)\Big|_{\vec a=(0, 0, a)}=-\frac{G M E^2}{2 |\vec p\, |}\sum_{\pm}(1\pm v)^2\log \mu(b\mp a)\ ,
\ee
which can be generalized to \eqref{delta_in_LogSum} by noticing that
\be
b\mp a=|\vec b\pm \hat p \times \vec a|\Big|_{\vec a=(0, 0, a)}\ .
\ee

For the KN case, the eikonal phase in the equatorial limit reads
\be\label{chaarge_eik_exact_equatrial}
\Delta\delta^{(1)}(p, \vec b\, )\Big|_{\vec a=(0, 0, a)}=\frac{G \pi Q^2}{4 a^2 |\vec p\, |}\left(-\frac{(a E+b |\vec p\, |)^2}{\sqrt{b^2-a^2}}+2aE|\vec p\, |+|\vec p\,|^2b\right)\ .
\ee
Taking the derivative with respect to $b$ one gets the scattering angle which coincides with the one in Eq. (17) of \cite{Hoogeveen:2023bqa}, modulo different conventions for the sign of the spin.
Moreover, we conjecture that generalizing \eqref{chaarge_eik_exact_equatrial} to arbitrary relative spin orientations should allow to write down a closed-form expression for this contribution to the eikonal phase.
Nevertheless we can expand \eqref{Eik_hFT_Explicit} in powers of $a$ and express the eikonal phase order by order in the impact parameter. This expansion has no subtleties and can be performed up to very high orders in $1/b$. For instance, the first three orders read 
\begin{equation}
    \Delta \delta^{(1)}\Big|_{1/b}=-\frac{G \pi Q^2 E}{8vb}(2+v^2)\ ,
\end{equation}
\begin{equation}
    \Delta \delta^{(1)}\Big|_{1/b^2}=-\frac{G \pi Q^2 a E \cos \beta}{4 b^2}\ ,
\end{equation}
\begin{equation}
    \Delta \delta^{(1)}\Big|_{1/b^3}=-\frac{G \pi Q^2 a^2 E}{128 v b^3}\Big(8+5v^2+(8+7v^2)\cos 2\beta+4(2+v^2)\cos 2\alpha \sin^2\beta\Big) \ ,
\end{equation}
from which following eq. \eqref{Deflection_Ang} we can easily write down the corresponding first three orders of the deflection angle 
\begin{equation}
    \Delta\vartheta^{(1)}\Big|_{1/b^2}=-\frac{G \pi Q^2}{4v^2b^2}(2+v^2)\ ,
\end{equation}
\begin{equation}
    \Delta \vartheta^{(1)}\Big|_{1/b^3}=-\frac{G \pi Q^2 a  \cos \beta}{v b^3}\ ,
\end{equation}
\begin{equation}
    \Delta \vartheta^{(1)}\Big|_{1/b^4}=-\frac{3 G \pi Q^2 a^2 }{64 v^2 b^4}\Big(8+5v^2+(8+7v^2)\cos 2\beta+4(2+v^2)\cos 2\alpha \sin^2\beta\Big) \ ,
\end{equation}
where we again make use of the splitting $\vartheta_{KN}=\vartheta+\Delta \vartheta$. Let us comment on this result. First we notice that the leading order contribution is $O(1/b^2)$, which means that the electromagnetic KN contribution to the deflection angle of a neutral probe is always subleading with respect to the Kerr one. Moreover, the leading order corresponds to the exact 1PM result for the non-rotating case, which in turn corresponds to a Reissner-Nordstr\"om BH. Finally since the leading order is negative, the overall sign of $\Delta \vartheta$ will be always negative, meaning that the KN contribution has the net effect to decrease the deflection angle. This would have been seen from eq. \eqref{gravitationalpotential}, that shows a relative minus sign between the two contributions, meaning that the charge of the BH has the effect of `decreasing' the curvature. It is also important to remark that once again the deflection angle depends only on the velocity of the probe, since even if we are considering the charge of the BH this is still a gravitational process, and the argument that holds in the Kerr case holds in the KN case as well. 

\begin{figure}[h]
\centering
\includegraphics[width=0.48\textwidth, valign=c]{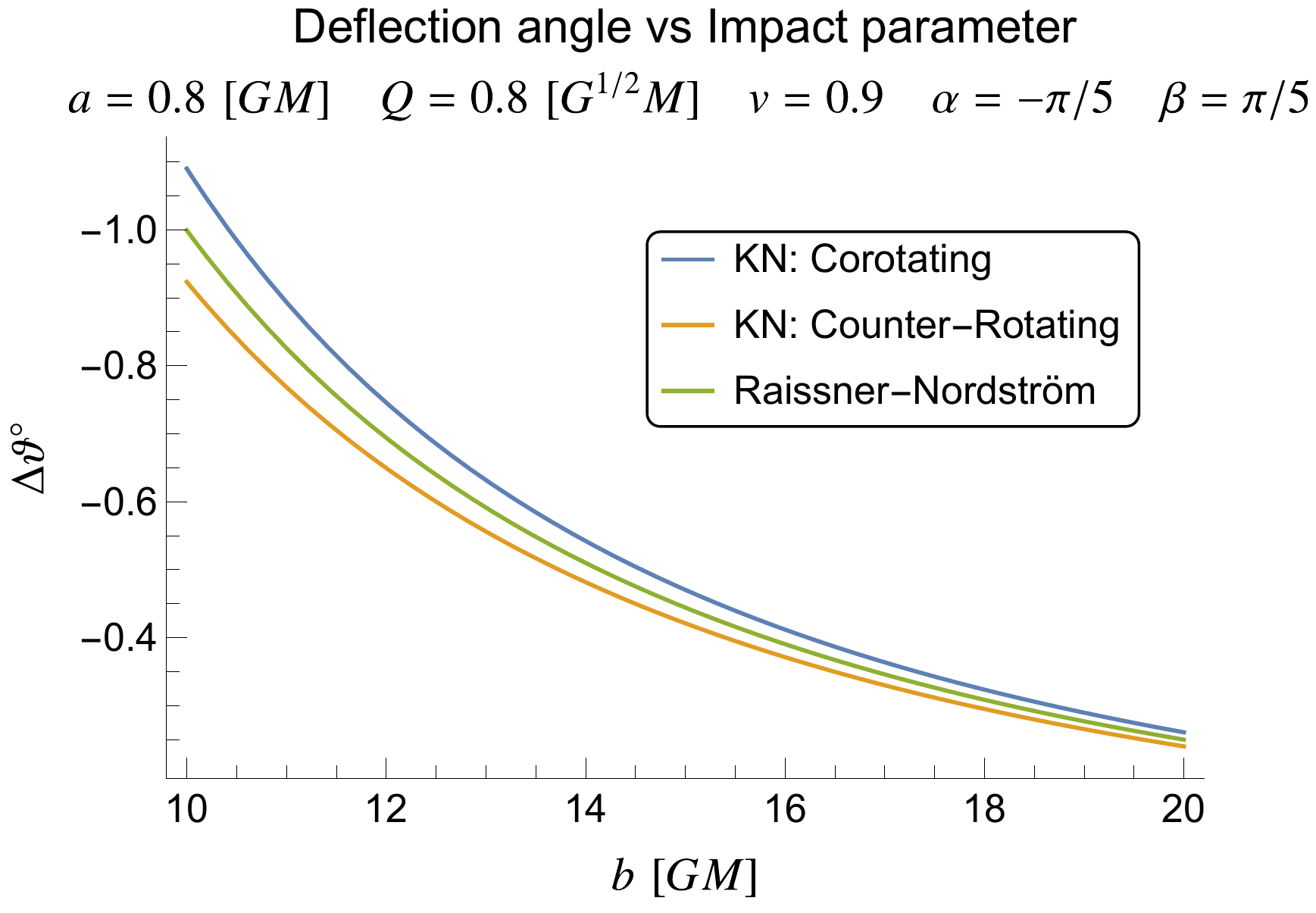}
\includegraphics[width=0.48\textwidth, valign=c]{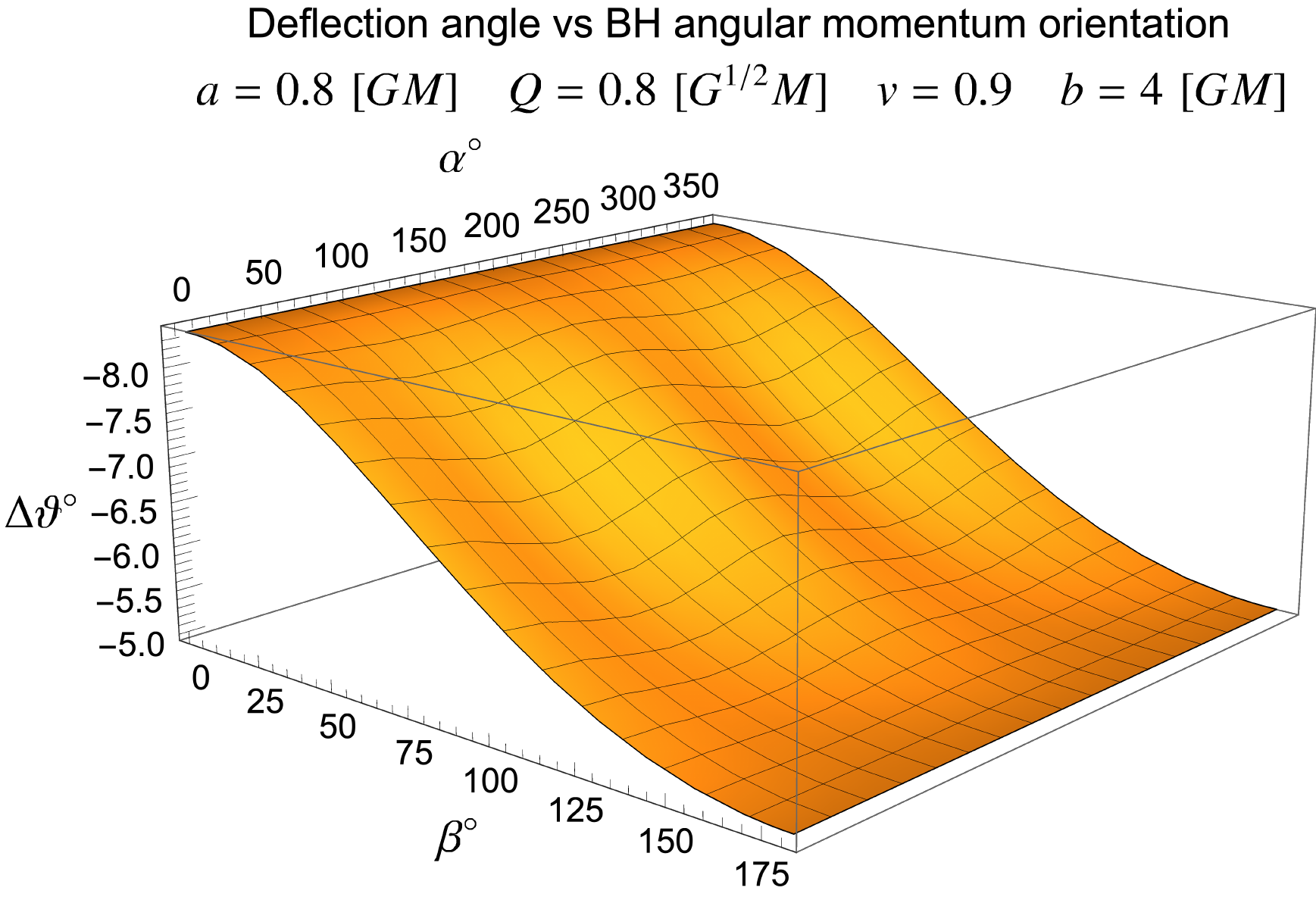}
\caption{Study of the deflection angle at 1PM for KN BHs. On the left the comparison of the deflection angle vs the impact parameter, with $v$, $Q$ and $\vec a$ fixed. On the right the deflection angle vs the orientation of the BH angular momentum for fixed $v$, $b$, $Q$ and $a$. Notice that these plots are obtained resumming the expansion of $\Delta\vartheta$ in powers of $1/b$ up to  the fifth order.}
\label{DeflectionAng_KN_Plot}
\end{figure}

We conclude this section by displaying some plots of the deflection angle in the KN case in Fig. \ref{DeflectionAng_KN_Plot}.
The plots show a very similar behaviour of the deflection angle for KN BHs as for Kerr BHs, and make it also manifest how in the same condition $\Delta\vartheta$ is a small correction of $\vartheta$.

\subsection{Gauge potential contribution}

A charged probe will interact also with the gauge potential generated by the KN  BH. The resulting contribution must be included in the evaluation of the total eikonal phase. At 1PM we have to consider the on-shell version of \eqref{MAamplitude}, which reads
\begin{equation}
    i\mathcal{M}_{on-shell}^A(p, \vec q\, )=-i \frac{4 \pi Q Q_\phi}{|\vec q\, |^2}\Bigg(2 E \cos|\vec a \times \vec q\, |+2i \sin|\vec a \times \vec q\, |\frac{\vec a\times \vec q\cdot\vec p}{|\vec a\times\vec q\, |}\Bigg)\ .
\end{equation}
For this amplitude exactly the same argument applies as for the Kerr case. Performing the replacement \eqref{LocalTerms_Simplification} and using eqs. \eqref{Cos_IPS} and \eqref{Sin_IPS}, one can express the eikonal phase in the compact formula
\begin{equation}
    \delta_A(p, \vec b\, )=\frac{Q Q_\phi}{2 v}\sum_{\pm}(1\pm v)\log|\vec b \pm \hat p\times \vec a|\ ,
\end{equation}
from which in turn one gets the deflection angle 
\begin{equation}\label{vartheta_GaugePotential}
    \vartheta_A(p, \vec b\, )=-\frac{Q Q_\phi}{v^2 E}\sum_{\pm}\frac{(1\pm v)(b\mp a\cos \beta)}{a^2\sin^2\alpha \sin^2\beta+(b\mp a\cos\beta)^2} \ .
\end{equation}
For scattering in the equatorial plane ($\beta=\pi$) this result can be compared with the term with $(n, j, k)=(0, 1, 0)$ of table IIIB of \cite{Hoogeveen:2023bqa}, showing perfect agreement.
In general, up to overall factors, this expression looks very similar to eq. \eqref{theta_Kerrcase}, in which now in the sum over the intermediate polarization appears $\sqrt{(1\pm v)^2}$. We know in fact that gauge amplitudes are square roots of gravity amplitudes \cite{Monteiro:2014cda,Kawai:1985xq}. In KS gauge this is classically reflected on the very expression of the gravitational field that is the `square' of the gauge potential $h_{\mu \nu}\sim \Phi^{-1} A_\mu A_\nu$, where $\Phi^{-1}$ eliminates spurious `double poles' \cite{Monteiro:2014cda}. 

Moreover, eq. \eqref{vartheta_GaugePotential} shows that if the charges of the BH and the probe have a different sign, the process is attractive exactly as one would expect. It is also important to notice that differently from the previous cases, the process now has an explicit dependence on the energy. This is due to the fact that this is not a gravitational process, and the dimensionless expansion parameter is just the electric charge. Finally we would like to compare the contribution to the deflection angle due to the gauge potential and the one due to gravity. If we focus on order of magnitude estimates for \eqref{theta_Kerrcase} and \eqref{vartheta_GaugePotential}, the two expressions will be comparable when 
\begin{equation}
    \frac{Q Q_\phi}{E b}\sim \frac{G M}{b} \ ,
\end{equation}
and since the typical BH charge is $Q\sim \sqrt{G}M$, in order to have $\vartheta_A\sim \vartheta$  one should take 
\begin{equation}
    Q_\phi=\frac{E}{M_p} \ ,
\end{equation}
where we have introduced the Planck mass $M_p\sim 1/\sqrt{G}$. Now if the typical charge of the probe is of order $Q_\phi\sim 1$ (for example in the electromagnetic case $Q_\phi\sim e \sim 1/\sqrt{137}$), then the two contribution will be of the same order of magnitude only if $E\sim M_p$, which is a huge energy. This means that for reasonable energies, the gauge contribution will be dominant over the gravitational interaction, exactly as one would expect.  

\section{Subleading eikonal corrections}\label{subleadingeikonalsection}

Let us discuss now the main feature of our KS approach, namely how higher orders in the PM expansion, including classical terms, arise only from the comb-like diagram in Fig. \ref{L_loop_Amp}.
Since the hyper-classical terms have already been discussed in section \ref{Sec:Eik_Expansion}, we will focus only on the classical contributions of eq. \eqref{MKN_0extra}.
We first consider the Schwarzschild case, and show how our method reproduces known results, and we then briefly discuss how to extend the analysis to KN. We also comment on how the graviton self-interaction contribution can be neglected in the probe limit approximation.

\subsection{2PM Schwarzschild classical term }

As already discussed in section \ref{Sec:Eik_Expansion}, the 2PM classical eikonal phase is obtained from the terms of the amplitude in \eqref{Amp_2loops} which are of order $O(1/\hbar^3)$. From the decomposition in eq. \eqref{M0_Mextra}, we obtain the amplitude written like \eqref{MKN_0extra}, from which the computation is reduced to evaluate those three pieces. For simplicity reasons we focus on the case in which the fixed background is a Schwarzschild BH, which means that from now on $\mathcal{M}_0$ and $\mathcal{M}_{extra}$ are respectively defined in eqs. \eqref{Mamplitude0} and \eqref{Mamplitudeextra} in the limit in which $a\rightarrow 0$. 

We start by discussing the $(0, 0)$ term referring to eq. \eqref{MKN_0extra}.
Considering the proper expansion of the propagator that follows from the $\hbar$ replacement rules,
\be
\frac{1}{(p-\hbar k)^2-m^2+i\varepsilon}=\frac{1}{\hbar \, \vec \ell\cdot \vec k +i\varepsilon}+\frac{|\vec k\, |^2- \vec k\cdot\vec q}{(\vec k\cdot \vec \ell\, )^2}+O(\hbar)\ ,
\ee
we select the classical contribution as
\be\label{00_Explicit_Amp}
\mathcal{M}^{(2)}\Big|^{(0, 0)}_{cl.}=8G^2M^2\pi^2(3|\vec p\, |^2+4 E^2)\int\frac{d^3k}{(2\pi)^3}\frac{1}{|\vec k\, |^2|\vec k -\vec q\, |^2}\ ,
\ee
where we have made use of the \textit{LiteRed} package \cite{LiteRed} to reduce the loop integrals. Finally, computing the master integral \cite{Mougiakakos:2020laz,DOnofrio:2022cvn}
\be
\int\frac{d^3k}{(2\pi)^3}\frac{1}{|\vec k\, |^2|\vec k -\vec q\, |^2} =\frac{1}{8}\frac{1}{|\vec q\, |}\ ,
\ee
we end up with 
\be\label{M_00_explicit}
\mathcal{M}^{(2)}\Big|^{(0, 0)}_{cl.}=G^2M^2\pi^2(3|\vec p\, |^2+4 E^2)\frac{1}{|\vec q\, |}\ .
\ee

Let us now discuss the $(extra, extra)$ term of eq. \eqref{MKN_0extra}. We can think of this term as the characteristic contribution due to the KS gauge, which is in fact composed from $i\mathcal{M}_{extra}$'s, that are the terms of the amplitude that are vanishing on-shell, and so hidden unless an off-shell formalism is used. 
As shown in section \ref{Sec:Eik_Expansion}, these $extra$ terms mimic the contact interactions in terms of comb-like diagrams (see eq. \eqref{Eq:ContactTerm}). 
In fact it is easy to see that $\mathcal{M}_{extra}$ exactly cancels the propagator as follows\footnote{Notice that we could cancel the propagator with $i\mathcal{M}_{extra}(p-k, p-q, \vec q-\vec k)$ as well.}
\be\label{scalar_contact}
i\mathcal{M}_{extra}(p, p-k, \vec k)\frac{i}{(p-k)^2-m^2+i\varepsilon}=i\frac{4 G M \pi^2 E}{|\vec k\, |^3}\ .
\ee
We can then write the $(extra, extra)$ contribution as
\be
\mathcal{M}^{(2)}\Big|^{(extra, extra)}_{cl.}=-16 G^2M^2\pi^4E^2\int\frac{d^3 k}{(2\pi)^3}\frac{\vec k\cdot (\vec q-\vec k\, )}{|\vec k\, |^3|\vec q- \vec k|^3}\ ,
\ee
where no \textit{LiteRed} identities are needed. Computing the master integral
\be
\begin{aligned}
\int\frac{d^3 k}{(2\pi)^3}\frac{\vec k\cdot (\vec q-\vec k\, )}{|\vec k\, |^3|\vec q- \vec k|^3}&\xrightarrow{\text{FT}}-\left(\partial_{\vec x}\mathcal{F}(3, -3/2) \right)\cdot\left(\partial_{\vec x}\mathcal{F}(3, -3/2) \right)=-\frac{1}{4\pi^4 |\vec x\, |^2}\\
&\xrightarrow{\text{FT}^{-1}}-\frac{1}{2\pi^2}\frac{1}{|\vec q\, |} \ , 
\end{aligned}
\ee
we can finally write the contribution 
\be\label{M_extraextra_Explicit}
\mathcal{M}^{(2)}\Big|^{(extra, extra)}_{cl.}=8 G^2M^2\pi^2E^2\frac{1}{|\vec q\, |}\ .
\ee

At last we consider the mixed term $(extra, 0)$. Computing it explicitly, one ends up with
\be\label{M_0extra_explicit}
2\mathcal{M}^{(2)}\Big|^{(extra, 0)}_{cl.}=i\,  G^2M^2\pi^3E\int\frac{d^3k}{(2\pi)^3} \vec k\cdot \vec p\ I(\vec k, \vec q\, )\ ,
\ee
where $I$ is independent of $\vec p$.
Without any further calculation, we can conclude that this term is vanishing. In fact, considering the tensorial structure of the integrand, we would have
\be
\int\frac{d^3k}{(2\pi)^3} \vec k\cdot \vec p\ I(\vec k,\vec q\, )= \vec q\cdot \vec p\ \tilde{I}_1(\vec q\, )=0\ ,
\ee
where $\tilde{I}$ is a function only of $\vec q$. As already pointed out in section \ref{Sec:Eik_Expansion}, since \eqref{M_0extra_explicit} is a pure imaginary contribution, this cancellation is not surprising. In fact we know that imaginary contributions are associated to radiative corrections, which however do not enter up to 3PM \cite{Damour:2020tta}. 

All in all, using eqs. \eqref{M_00_explicit} and \eqref{M_extraextra_Explicit}, we can finally write the eikonal phase at 2PM in the Schwarzschild case as 
\be
\begin{aligned}
\delta^{(2)}(p, \vec b\,)\Big|_{a=0}&=\frac{1}{4|\vec p\, |}\int\frac{d^2q}{(2\pi)^2}e^{i\vec q \cdot \vec b}\left(\mathcal{M}^{(2)}\Big|^{(0,0)}_{cl.}+\mathcal{M}^{(2)}\Big|^{(extra, extra)}_{cl.}\right)\\
&=\frac{3G^2M^2\pi E}{8vb}(4+v^2)\ ,
\end{aligned}
\ee
reproducing the results in \cite{Damour:2017zjx, KoemansCollado:2018hss}. As usual, we can express the deflection angle obtaining the well known expression
\be
\vartheta^{(2)}\Big|_{a=0}=\frac{3G^2M^2\pi^2}{4v^2b^2}(4+v^2)\ .
\ee

\subsection{2PM Kerr classical term}

With more effort one can extend the analysis to rotating BHs. As in the case of Schwarzschild, working in the KS gauge it is convenient to separate the building block tree-level amplitude given in eq. \eqref{M0_Mextra} in two parts, namely $i\mathcal{M}_0$ and $i\mathcal{M}_{extra}$, defined respectively in eqs. \eqref{Mamplitude0} and \eqref{Mamplitudeextra}.\footnote{Here the discussion is  restricted to Kerr BHs, but can be generalized to KN BHs as well.}
Anyway, after plugging these terms in eq. \eqref{Amp_2loops}, the actual computation of the integrals seems rather involved and therefore, despite the conceptual simplicity of this KS approach, finding an explicit expression for $\delta^{(2)}$ is beyond the scope of the present investigation.  Nonetheless the very form of the integrals suggests the observations that follow.

As discussed in section \ref{Sec:Eik_Expansion}, referring to eq. \eqref{Amp_2loops}, the hyper-classical term is reproduced by considering those term in $i\mathcal{M}^{(2)}$ of order $O(1/\hbar^4)$ which is easily obtained by a convolution of 1PM amplitudes, so that we only care about the classical term. This one follows the structure of \eqref{MKN_0extra}, and as in the Schwarzschild case we need to compute three different terms involving integrals of different kind. 
The term $(0, 0)$ involves integrals of the form  
\be
i\mathcal{M}^{(2)}\Big|_{(0, 0)}^{cl.}=G^2M^2\int\frac{d^3 k}{(2\pi)^3} \frac{j_{a}(|\vec a\times\vec k|)j_b(|\vec a\times(\vec q-\vec k)|)}{|\vec k|^{m}|\vec q-\vec k|^{n}}I^{(0,0)}_{a,b, m, n}(\vec k, \vec q, \vec \ell, \vec a\, ) \ ,
\ee
where here and in the following $a, b=0, 1$, $m, n=0, \dots, 4$ and $I^{(0,0)}_{a,b, m, n}(\vec k, \vec q, \vec \ell, \vec a\, )$ are rational functions of vector and scalar products of the arguments.  
Then, the term $(extra, 0)$ has the structure
\be
2i\mathcal{M}^{(2)}\Big|_{(extra, 0)}^{cl.}=G^2M^2\int\frac{d^3 k}{(2\pi)^3} \frac{J_{a}(|\vec a\times\vec k|)j_b(|\vec a\times(\vec q-\vec k)|)}{|\vec k|^{m}|\vec q-\vec k|^{n}}I^{({ex},0)}_{a,b, m, n}(\vec k, \vec q, \vec \ell, \vec a\, ) \ ,
\ee
and finally the term $(extra, extra)$ reads
\be
i\mathcal{M}^{(2)}\Big|_{(extra, extra)}^{cl.}=G^2M^2\int\frac{d^3 k}{(2\pi)^3} \frac{J_{a}(|\vec a\times\vec k|)J_b(|\vec a\times(\vec q-\vec k)|)}{|\vec k|^{m}|\vec q-\vec k|^{n}}I^{({ex},{ex})}_{a,b, m, n}(\vec k, \vec q, \vec \ell, \vec a\, ) \ .
\ee
As mentioned above, even if the systematic computation of such integrals is more laborious than in the non-rotating case we can still make some general statements. 

First of all, comparing with the Schwarzschild case discussed in the previous subsection, a property that still holds  in the Kerr case is the fact that $i\mathcal{M}_{extra}(p, p-k)$, as well as $i\mathcal{M}_{extra}(p-k, p-q)$, exactly cancel the massive propagator mimicking the contact terms. This cancellation explicitly reads
\be
\begin{gathered}
i\mathcal{M}_{extra}(p, p-k, \vec k\, )\frac{i}{(p-k)^2-m^2+i\varepsilon}\\
=\frac{2G M \pi^2}{|\vec a\times \vec k\, ||\vec k\, |^3}\Big(2 i E |\vec a\times \vec k\, |J_0(|\vec a \times\vec k\, |)-J_1(|\vec a\times\vec k|)\vec a \times \vec k\cdot( \vec \ell+ \vec q)\Big)\ ,
\end{gathered}
\ee
where it is easy to see that for $a\rightarrow 0$ we reproduce eq. \eqref{scalar_contact}. Notice that for the amplitude $i\mathcal{M}_{extra}(p-k, p-q)$ the result is analogue. 

As we have seen in the previous subsection, another property of the Schwarzschild case is the vanishing of  the $(extra, 0)$ term. A natural guess would be that such term cancels out for rotating BHs as well. Encouragingly, one can prove that this term vanishes up to second order in the angular momentum of the BH, namely 
\be
2i\mathcal{M}^{(2)}\Big|_{(extra, 0)}^{cl.}=O(a^2)\ .
\ee
Extending the analysis to higher orders in $a/b$ is tedious and we leave it to the future to confirm our expectations.  

To conclude, even if at the moment we do not have an explicit expression for $\delta^{(2)}$, the KS gauge approach looks very promising in setting up the calculations order by order in the PM expansion and we plan to return soon to this and related issues. 

\subsection{Graviton self-interaction contribution}

At the very end, one may wonder whether diagrams generated by the exchange of virtual gravitons with non space-like momenta may contribute at sub-leading order already at 2PM. Expanding the metric around a background in the KS gauge as
\be
\bar{g}_{\mu\nu}={g}_{\mu\nu}+\kappa \delta  h_{\mu\nu}=\eta_{\mu\nu}+h_{\mu\nu}+\kappa \delta h_{\mu\nu}\ ,
\ee
where $\delta h_{\mu\nu}$ is the graviton field, $h_{\mu\nu}$ is defined in eq. \eqref{KerrSchieldx}, and $\kappa^2=32\pi G$,  the Einstein-Hilbert action gives the interaction vertices among the gravitons themselves, as well as mixed vertices in which the gravitons interact with the source. Since the background metric satisfies the Einstein equations, one can show that the tri-linear vertex in which two sources interact with one graviton vanishes, namely
\be
V_{\delta h, h, h} =0\ ,
\ee
as well as all the vertices $V_{\delta h, h,..., h}=0$ with an arbitrary number of sources \cite{Donoghue:1995cz}. Clearly for Schwarzschild and for Kerr this is enough, while for Kerr-Newman one has to take into account also the gauge potential, which is expanded around the external source as
\be
\bar{A}_\mu=A_\mu+\delta A_\mu\ ,
\ee
where again $\delta A_\mu$ is the quantum fluctuation (the photon) and $A_\mu$ is the classical background in the KS gauge given in eq. \eqref{KerrSchieldAx}. Given that also the gauge potential $A_\mu$ satisfies the field equations, then the correct expression is
\be
V_{\delta h, h, h} + V_{\delta h, A, A} =0 \ ,
\ee
as well as 
\be
V_{\delta A, A, h} =0
\ee
for charged particles.

\begin{figure}[h]
\centering
\includegraphics[width=0.4\textwidth, valign=c]{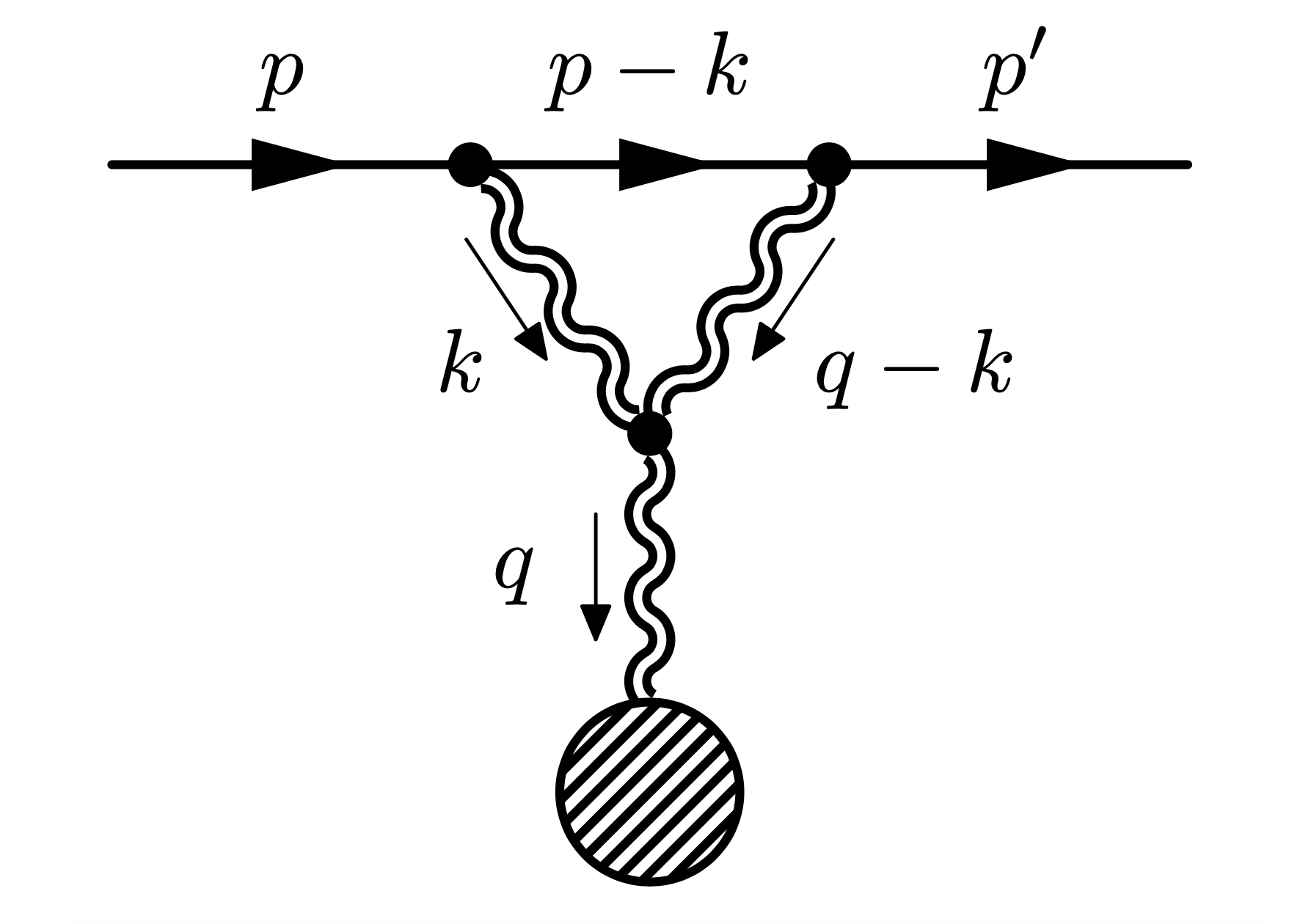}
\caption{Diagram in which the probe interacts with the source through a loop of virtual gravitons.}
\label{3Grav_Diagram}
\end{figure}

However, at 2PM there is another vertex that one has to consider, which is $V_{\delta h, \delta h, h}$, leading to the diagram in Fig. \ref{3Grav_Diagram}.
In order to compute such diagram, 
we consider the vertex $V_{\delta h^, \delta h, h}$ in \cite{Donoghue:1995cz} and the 2 scalar - 1 graviton vertex $V_\phi$,  defined as 
\be
V_\phi^{\mu\nu}(p, p')=- \frac{i}{2} \kappa \widetilde{T}^\phi_{\mu\nu} (p, p' )\ ,
\ee
so that the amplitude  reads
\be
\begin{aligned}
i\mathcal{M}^{(2)}\Big|_{\delta h, \delta h, h}=\int\frac{d^4k}{(2\pi)^4}&\frac{iV_\phi(p, p-k)V_\phi(p-k, p-q)}{(p-k)^2-m^2+i\varepsilon}V_{\delta h, \delta h, h}(k, q-k)\\
&\times\frac{i P}{k^2+i\varepsilon}\frac{i P}{(q-k)^2+i\varepsilon}i \kappa^{-1} \tilde{h}^{KN}(\vec q\,) \ ,
\end{aligned}
\ee
where $P$ is the tensorial structure of the graviton propagator and all the contractions are left understood. 
Now we want to extract out of this amplitude the classical term,  that as discussed in section \ref{Sec:Eik_Expansion} is the term
\be
i\mathcal{M}^{(2)}\Big|_{\delta h, \delta h, h}^{cl.}=O(1/\hbar^3)\ .
\ee
This imposes to put on-shell the massive legs and to reduce the scalar propagator to $\delta(\ell\cdot  k\, )$, from which we can integrate out the temporal component of the internal momentum by putting 
\be
k_0=\frac{\vec p\cdot \vec k}{E}\ .
\ee
All in all, the amplitude becomes
\be
i\mathcal{M}^{(2)}\Big|^{cl.}_{\delta h, \delta h, h}=-i\frac{1}{4E\kappa}\int\frac{d^3k}{(2\pi)^3}\frac{V_\phi(p,p)V_\phi(p,p)V_{\delta h, \delta h, h}(k, q-k)\tilde{h}^{KN}(\vec q\, )\ PP}{\left((\frac{\vec p\cdot \vec k}{E})^2-|\vec k\, |^2\right)\left((\frac{\vec p\cdot \vec k}{E})^2-|\vec q-\vec k\, |^2\right) }\ .
\ee
The explicit computation of the integral is not needed, since by looking at the structure of the amplitude, we can immediately see that it is of order $O(G^2M)$. This means that in the probe limit in which $M\rightarrow +\infty$, this contribution is totally negligible with respect to the classical terms computed at the beginning of the section. Notice that this could have already  been deduced by simply looking at Fig. \ref{3Grav_Diagram}.

Finally we notice that in principle one should consider two more diagrams similar to the one in Fig. \ref{3Grav_Diagram}, namely the ones involving $V_{\delta A, \delta A, h}$ and $V_{\delta h, \delta A, A}$, with photons running in the loops. The exact same argument discussed before holds also in this case, and all these diagrams are suppressed by factors $E/M\ll1$ and $Q_\phi/Q\ll1$.
We then conclude by saying that in the probe limit at 2PM the only terms that contribute to $\delta^{(2)}$ are the classical contributions coming from the comb-like diagram. 

\section{Conclusions and outlook}\label{Sec:Conclusions}

Let us conclude with a summary of our results and draw lines for future investigation.
Exploiting the Kerr-Schild gauge, we have derived a very compact analytic form for the scattering amplitude of a massive charged scalar off a KN BH. The remarkable feature of this gauge is the presence of a single tri-linear vertex and no higher order interaction vertices. Putting the external legs on-shell has allowed us to write down a very compact Rutherford-like formula for the elastic cross-section with arbitrary relative orientation of the spins of the probe and of the black hole. 
We have then shown that terms that are zero on-shell actually contribute at higher order to effectively generate higher point `contact' interactions. This has been explicitly shown at order 2PM for scattering off a Schwarzschild BH, while for rotating BHs we have presented a schematic analysis and plan to give a more detailed derivation in the near future.

The charge plays a dual role in the game. On the one hand it contributes to the energy of the background that corrects the dynamics even of neutral probes, and  on the other hand, for charged probes, it produces new terms due to `photon' exchange. 
These terms are dominant for elementary particle probes, but can also be relevant for BH-BH interactions,  like in the inspiral phase of mergers, involving near extremal BHs.
Although electrically charged BHs are expected to discharge quite rapidly, this is not the case for BHs carrying `dark-photon' charges \cite{Cardoso:2016olt}. In fact dynamics of mini-charged probes can set bounds on dark matter candidates. In this sense our analysis can prove to have some phenomenological implications.

It would be nice to extend our study in different directions. First of all, as already mentioned it would be very interesting to extend the 2PM analysis to the KN case. In the Kerr case, while in general this problem is tackled using higher-spin amplitude methods, in the work of \cite{Bautista:2022wjf, Bautista:2023szu} the 2PM result is exact in the spin of the black hole, thanks to an ansatz for  the gravitational Compton amplitude. Our method should reproduce these results, as well as classical results obtained in \cite{Damgaard:2022jem, Gonzo:2023goe}, and also give a natural way to extend them to higher PM orders. Indeed, computational difficulties aside, our approach circumvents the problem of determining an $n$-point vertex which is exact in the spin. 
Moreover, it would also be relevant to extend the analysis to probes with spin. As we have shown, in the case of spin 1 there are no conceptual differences with respect to the scalar case, and in particular for massless and chargeless massive vectors only a tri-linear vertex is present, while in the case of a charged massive vectors there is also a quartic coupling with the background gauge potential. The generalization to a spin 2 probe, relevant for the study of the scattering of gravitational waves, is left as an open problem.
Finally, we would like to understand how to generalize our analysis beyond the probe approximation to obtain results for the 2-2 scattering of spinning black holes.

There are also other directions that can be pursued.
One  could exploit the KS gauge for other backgrounds such as AdS, AdS BHs and (rotating) BHs and branes in higher dimensions \cite{Monteiro:2014cda}. We hope to report soon on some of these issues.
Moreover, the leading eikonal phase can be compared with the results of the partial wave expansion computed in the wave scattering approach based on BHPT \cite{Bautista:2021wfy, Bautista:2022wjf}. The relevant wave equation, known as Teukolsky equation \cite{Teukolsky:1973ha,Press:1973zz,Teukolsky:1974yv}, separates into a radial and a polar angular part. Both can be put in the form of a Confluent Heun Equation with two regular singularities (at the horizons) and an irregular singularity at infinity \cite{Doran:2001ag, Glampedakis:2001cx,Dolan:2008kf}. Quite remarkably, the very same equations  can be regarded as quantum Seiberg-Witten curves for ${\cal N}=2$ SYM theory with $SU(2)$ gauge group and $N_f=(2,1)$ flavours (hypermultiplets) in the fundamental (doublet) representation \cite{Aminov:2020yma, Bianchi:2021xpr, Bonelli:2021uvf, Bianchi:2021mft, Bonelli:2022ten, Consoli:2022eey, Bianchi:2022qph, Bianchi:2023rlt, Bianchi:2023sfs}. This approach, combined with the AGT correspondence~\cite{Alday:2009aq}, proves to be very effective in the determination of the relevant connection formulae and in the computation of the spectrum of Quasi Normal Modes, absorption probability, super-radiance amplification factor and Tidal Love Numbers. In all cases one imposes ingoing boundary conditions at the event horizon. Direct comparison should not be limited to the case of incidence along the $z$ axis for which results for the phase shift are already available in BHPT approach \cite{Doran:2001ag, Glampedakis:2001cx,Dolan:2008kf} but, using the above approach, may apply to arbitrary relative spin cases. 

Further simplification occurs for large $b$, for which the deflection angle is small and can be estimated by saddle-point approximation and related to the action for geodetic motion in the KN background
\be
S= \int P_r (J_z,E, K; r) dr + \int P_\theta(J_z,E, K; \theta) d\theta\ ,
\ee
where $K$ is Carter's separation constant. Both can be written as (in)complete elliptic integrals. Quite remarkably in the extremal case for $a=M$, thanks to generalized Couch-Torrence conformal inversions \cite{Couch1984ConformalIU, Cvetic:2020kwf, Cvetic:2021lss, Bianchi:2021yqs, Bianchi:2022wku}, the radial action is invariant under 
\be
r-a\rightarrow \frac{(r_c - a)^2}{r-a}\ ,
\ee
that exchanges the horizon with infinity keeping fixed the photon-sphere $r=r_c$ as well as the polar angular action. This should reflect in special properties of the scattering amplitudes, eikonal phases, and ultimately of the cross-section that are awaiting to be discovered.

\vskip 1cm

\section*{Acknowledgments}

We would like to thank Y.F. Bautista, D. Bini, A. Brandhuber, T. Damour, M. Firrotta, A. Guevara, Y.-T. Huang, C. Kavanagh, J.-W. Kim, M. Marzi, J.F. Morales, P. Pani, R. Russo, A. Salvio, F. Tombesi, G. Travaglini and J. Vines for discussions and comments on the manuscript.
We thank the GGI, where part of this work was carried out, for hospitality during the Mini Workshop ``New horizons for horizonless physics: from gauge to gravity and back again''.
Our work is partially supported by the MIUR PRIN Grant 2020KR4KN2 ``String
Theory as a bridge between Gauge Theories and Quantum Gravity''.

\vskip 1cm

\bibliographystyle{JHEP}
\bibliography{biblio}

\end{document}